\documentclass[useAMS,usenatbib]{mn2e}
\usepackage{latexsym,hhline,epsfig,longtable}
\usepackage{graphicx,color}
\usepackage{ulem}

\def\chem#1{$^{#1}$}                                           
 
\title[Low metallicity super-AGB star yields]{Super and massive AGB stars - III. Nucleosynthesis in metal-poor and very metal-poor stars - Z=0.001 and 0.0001}
\author[C.L. Doherty, P.Gil-Pons, H.H.B Lau, J.C Lattanzio, L. Siess, S.W. Campbell]{Carolyn L. Doherty$^{1}$\thanks{E-mail:carolyn.doherty@monash.edu}, Pilar Gil-Pons$^{2}$, Herbert H.B. Lau$^{3}$, John C. Lattanzio$^{1}$,
\newauthor  Lionel Siess$^{4}$ and Simon W. Campbell$^{1}$ \\
$^{1}$Monash Centre for Astrophysics (MoCA), School of 
Mathematical Sciences, Monash University, Victoria 3800, Australia\\
$^{2}$ Department of Applied Physics, Polytechnical University of Catalonia, 08860 Barcelona, Spain\\
$^{3}$Argelander Institute for Astronomy, University of Bonn, Auf dem Huegel 71, D-53121 Bonn, Germany\\
$^{4}$Institute d’Astronomie et d’Astrophysique, Universit Libre de Bruxelles, CP 226, B-1050 Brussels, Belgium} 

\begin{document}
\maketitle

\begin{abstract}

We present a new grid of stellar models and nucleosynthetic yields for super-AGB stars with metallicities Z=0.001 and 0.0001, applicable for use within galactic chemical evolution models. 
Contrary to more metal rich stars where hot bottom burning is the main driver of the surface composition, in these lower metallicity models the effect of third dredge-up and corrosive second dredge-up also have a strong impact on the yields.
These metal-poor and very metal-poor super-AGB stars create large amounts of \chem{4}He, \chem{13}C and \chem{14}N, as well as the heavy magnesium isotopes \chem{25}Mg and \chem{26}Mg. There is a transition in yield trends at metallicity Z$\approx$0.001, below which we find positive yields of \chem{12}C, \chem{16}O, \chem{15}N, \chem{27}Al and \chem{28}Si, which is not the case for higher metallicities.
We explore the large uncertainties derived from wind prescriptions in super-AGB stars, finding $\approx$ 2 orders of magnitude difference in yields of \chem{22}Ne, \chem{23}Na, \chem{24,25,26}Mg, \chem{27}Al and our s-process proxy isotope $g$. We find inclusion of variable composition low temperature molecular opacities is only critical for super-AGB stars of metallicities below Z$\approx$0.001.
We analyze our results, and those in the literature, to address the question: Are super-AGB stars the polluters responsible for extreme population in the globular cluster NGC 2808?  Our results, as well as those from previous studies, seem unable to satisfactorily match the extreme population in this globular cluster. 

\end{abstract}


\section{Introduction}

Super-AGB stars are characterised by off-centre carbon ignition and at the low metallicities studied in this work have initial masses between $\sim$ 6.5 and 9.0\,M$_\odot$. They undergo from tens to thousands of thermal pulses and associated third dredge-up (3DU) episodes which enrich their envelopes with the products of nuclear burning. They also have relatively extreme nucleosynthetic conditions with temperatures at the base of the convective envelope reaching over 130MK which leads to efficient hot bottom burning (HBB). 
Because of their short lifetimes (30-50 Myrs), metal-poor and very metal-poor (Z=0.001 and 0.0001) super-AGB stars are some of the first AGB stars to have enriched the interstellar medium. 
 
While nucleosynthesis in intermediate-mass and massive\footnote{We define massive AGB stars as those with initial masses $\ga$ 5\,M$_\odot$ but not massive enough to ignite carbon.} metal-poor and very metal-poor AGB stars has been explored in considerable detail \cite[e.g.][]{ven02,den03,her04a,fen04,ven05,mar09,ven09b,kar10,lug12a,dor13a}, to date, at these metallicities there have only been a few nucleosynthesis studies along the entire super-AGB phase \citep{sie10,ven10a,ven13}, with a notable absence of third dredge-up in these works.

At moderate metallicities ([Fe/H] $\ga$ $- 0.6$), massive AGB star models can be directly compared to observations to constrain the occurrence and relative impact of HBB and 3DU on the nucleosynthesis \cite[e.g.][]{ven00,mcs07,van12,gar13}. At the lower metallicities considered here however, super-AGB and massive AGB stars will have died long ago, making direct comparison impossible.  
Nevertheless there is the possibility to derive some constraints on the evolution and nucleosynthesis of low-metallicity super-AGB and massive AGB star evolution indirectly via galactic chemical evolution models. For example super-AGB and massive AGB stars may have contributed to the rise of heavy magnesium isotopes in the Galaxy \citep{fen03} and may be an extra source of \chem{13}C and \chem{14}N in the early stages of the Galaxy's formation. 

Super-AGB stars may also play a role as polluters within globular clusters. The ``abundance anomaly problem'' in globular clusters (GCs) has been one of the most thoroughly researched subjects in modern astrophysics \cite[e.g.][]{kra79,kra94,gra12}. A substantial fraction of stars within GCs are found to have unusual compositions (not seen in field stars), characterised by the results of hot hydrogen burning \citep{den90}. These stars show variations in a number of  light elements such as C, N, O, F Na, Mg, Al and Si. However, the anomalous and normal stars typically have the same [Fe/H], total number abundance of C+N+O \cite[e.g.][]{smi96,iva99}\footnote{The constancy of total CNO abundance between anomalous and normal stars within GCs is debated, for example in NGC 1851 where it is found to vary by a factor of 2 \citep{yon09} or not at all \citep{vil10}.}, lithium \citep{dor10b} and s-process element abundances \cite[e.g.][]{yon06b,yon08a}. These shared traits place strong constraints on the source of the anomalous material.

One of the main theories to explain these abundance anomalies is that GCs are made of multiple generations of stars, with the anomalous stars being formed from the enriched material of a first generation of stars (for a review see \citealt{gra12}). Stars within GCs are also further divided into three main populations based on their O and Na abundances; primordial (P) which comprises of first generation stars and intermediate (I) and extreme (E) which belong to the second generation. Generally only the most massive CGs harbour an extreme population \citep{car09a} showing the largest abundance anomalies, with substantial depletion of C, O, Mg and large enhancement of He, N, Na and Al.

Although the source of gas from which the suspected second generation of stars formed is still uncertain, proposed candidates include: intermediate-mass (and super-)AGB stars \cite[]{cot81,ven01,der08}, winds from fast rotating massive stars \cite[]{nor04,dec07b}, massive star binaries \citep{dem09} and super-massive stars \citep{den14} to name a few. While each of these classes of polluter has their associated problems matching certain observed features \citep{gra12} here we focus on the super-AGB and massive AGB star scenario. 
\cite{der08,der12} have suggested the extreme population within GCs are formed directly from pristine super-AGB ejecta, with the intermediate population formed via some dilution of polluted material with pristine gas \cite[e.g.][]{bek07,dan07,der08,der10,gra10}. 

This paper is the third in this series dedicated to the study of super-AGB stars and is organized as follows: Section~\ref{sec-code} summarizes our numerical program and input physics, in Section~\ref{sec-nucleo} we explore the nucleosynthesis of these low metallicity models, in Section~\ref{yields} we discuss the stellar yields, and a range of uncertainties. Comparisons to the past works on super-AGB nucleosynthesis are made in Section~\ref{sec-ngc2808} and then we apply these yield results to the extreme population in the globular cluster NGC 2808. In Section~\ref{sec-conclude} we summarize and conclude.

\section{Stellar Evolution and Nucleosynthesis Programs}\label{sec-code}

The Monash University stellar evolution program (\textsc{monstar}; for details see \citealt{cam08,doh10}) was used to calculate the structural evolution models. Whilst this present work focuses on nucleosynthesis, these evolutionary models will be discussed in more detail in Paper IV (Doherty et al., in preparation) in this series. 
We briefly describe the relevant input physics. The \cite{rei75} mass-loss rate with $\eta$=1.0 is used prior to the carbon burning phase where we switch to the mass-loss rate from \cite{vas93}. For the mixing-length parameter we use $\alpha_{\rm{mlt}}$=1.75, calibrated to the standard solar model. We employ the search for convective neutrality as described in \cite{lat86} to determine the convective boundaries. High temperature opacities are from the OPAL compilation \citep{opal} whilst low temperature opacities are from \cite{fer05b}. In some specific test models, we also used the variable C, N composition molecular opacities from \cite{led09}.

Nucleosynthesis calculations were performed using the Monash University stellar nucleosynthesis post-processing program \textsc{monsoon} which solves simultaneously chemical transport and nuclear burning \cite[for details see][]{can93,lug04,doh14a}. The nuclear network comprises 77 species up to sulfur (including both ground and metastable states of \chem{26}Al) and includes the iron group elements, using the double sink approach as originally proposed by \cite{jor89}. To terminate the network and simulate neutron capture reactions beyond nickel, we use a sink particle defined as $g$, as our s-process proxy. The initial abundances were taken from the compilation of \cite{gre96} with scaled solar composition.

We calculate nucleosynthetic yields and provide online tables as in Paper II \citep{doh14a}. We present stellar yields for super-AGB stars of 6.5, 7.0 and 7.5\,M$_\odot$ for Z=0.001 and 0.0001. 

\section{Nucleosynthesis Results}\label{sec-nucleo}

\begin{figure*}
\begin{center}
\includegraphics[width=10.0cm,angle=0]{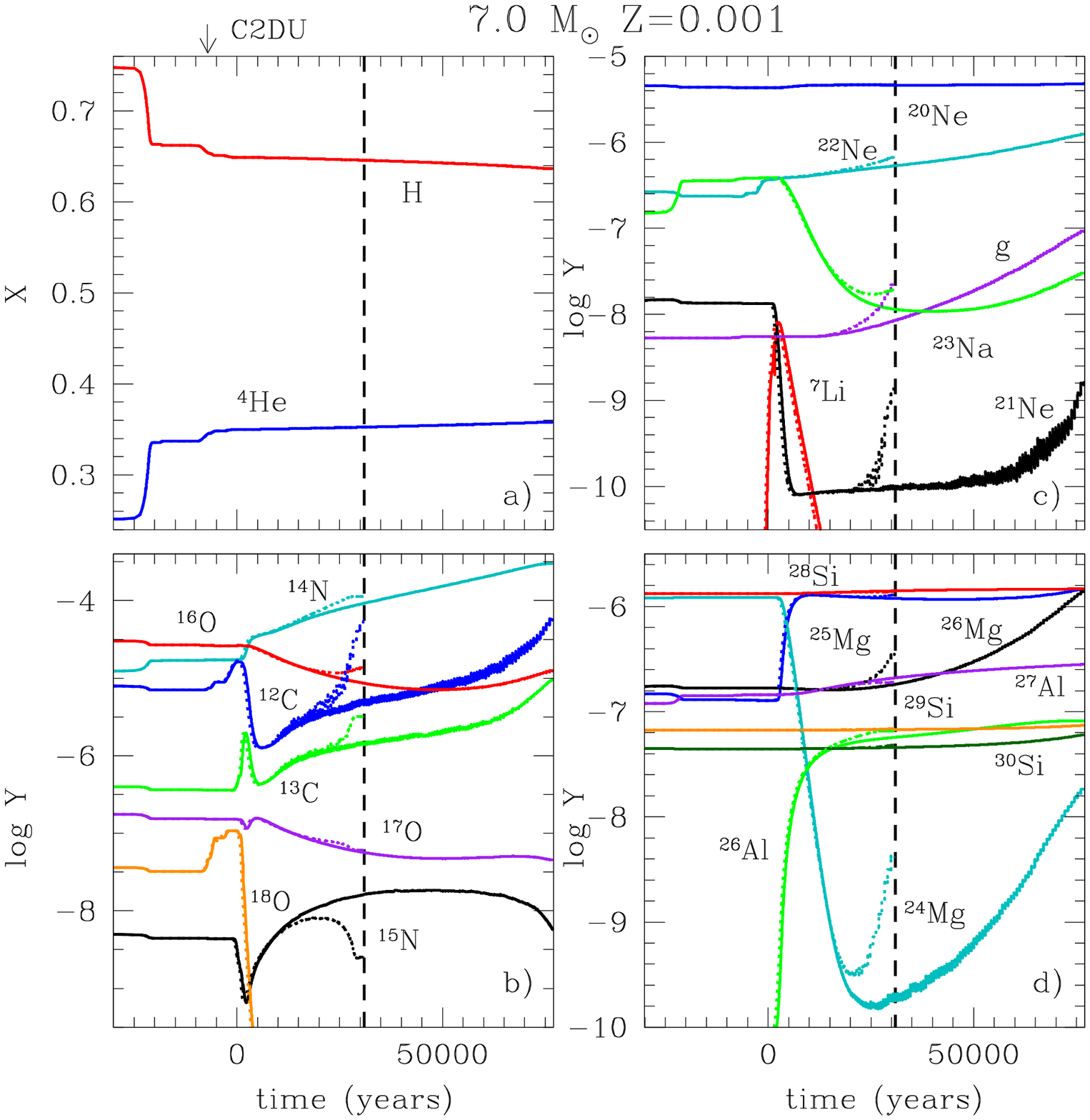}
\includegraphics[width=10.0cm,angle=0]{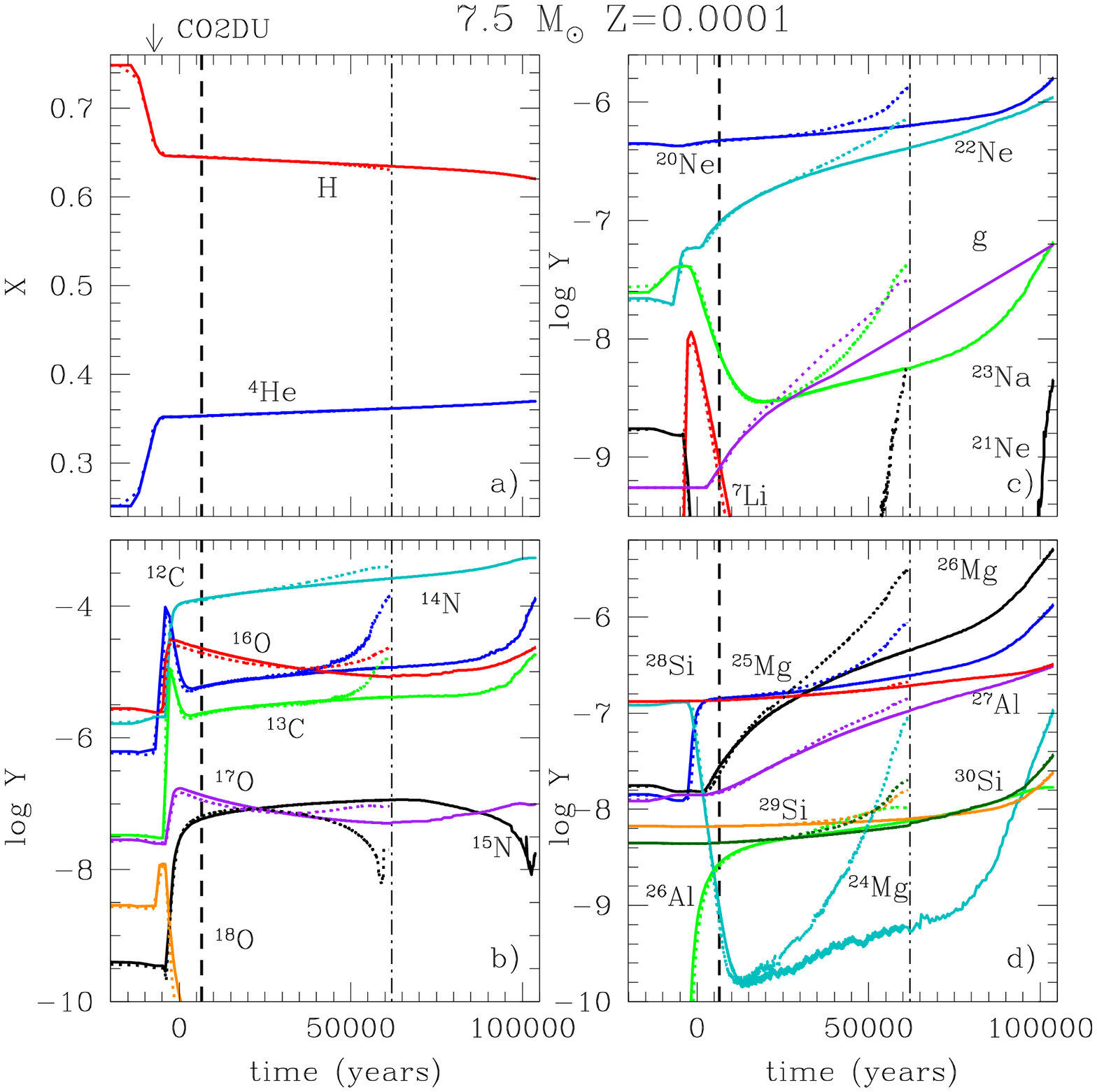}
\caption{\small{Surface abundances (in mass fraction for H and He otherwise in mole fraction Y) of selected isotopes as a function of time for the 7.0\,M$_\odot$ Z=0.001 (top) and 7.5\,M$_\odot$ Z=0.0001 (bottom) models.  Panels show a) H and He, b) CNO isotopes c) Ne-Na isotopes, \chem{7}Li and $g$ d) Mg-Al and Si isotopes. The time axis has been offset with the zero at the time of the first thermal pulse. In the top panel the results using the mass-loss rate of VW93 are shown with solid lines and the B95 mass-loss rate models are shown with dotted lines. In the bottom panel we show results for the VW93 model again with solid lines whilst the models with VW93 and variable composition Low-T molecular opacities are shown with dotted lines. The dashed vertical lines represent the truncated duration of the TP-(S)AGB phase for the B95 mass-loss rate, whilst the dot-dashed line shows the end of the evolution in the variable composition low-T opacity model. The Z=0.0001 B95 model is not shown for clarity, however, it does follow the VW93 abundance patterns closely. The onset of the C2DU and CO2DU events are also labeled.}}\label{fig-surf}
\end{center}
\end{figure*}

\subsection{Second Dredge-Up}\label{sec-2du}

Due to rapid ignition of core helium burning, low metallicity, intermediate-mass stars do not reach the first giant branch nor undergo a first dredge-up event \citep{gir96}. Therefore the first time the surface is enriched in nuclear processed material is the second dredge-up event (2DU). Traditionally, a 2DU event brings products that have undergone CNO cycling to the surface (primarily \chem{14}N). As the CNO cycle is generally a catalytic process with minimal leakage through to \chem{19}F or \chem{20}Ne, a standard 2DU results in a change in the relative proportions of C, N and O at the surface but \textit{no} net increase in C+N+O.
However, in our more massive super-AGB star models, during the 2DU the inward moving convective envelope penetrates deep and into the top of the helium burning shell. 
 The resultant nucleosynthesis from this corrosive second dredge-up (C2DU) leads to a large increase in the surface abundances of the products of partial or complete helium burning, namely \chem{12}C and \chem{16}O, as well as \chem{18}O and \chem{22}Ne synthesised via \chem{14}N($\alpha$,$\gamma$)\chem{18}F($\beta^+$$\nu$)\chem{18}O($\alpha$,$\gamma$)\chem{22}Ne. We subdivide corrosive 2DU into two types (which correlate with increasing core mass), either C2DU with only carbon enhancement or CO2DU with increases in both C and O. 

Fig.~\ref{fig-surf} illustrates  the evolution of the surface abundances prior to the first thermal pulse during the C2DU of the 7.0\,M$_\odot$ Z=0.001 model and the CO2DU event in the 7.5\,M$_\odot$ Z=0.0001 model.
The amount of He burning processed material dredged to the surface during one of these events is quite significant, with the number ratio of total C+N+O to initial C+N+O increasing to $\sim$ 1.2, 3.5 for the 7.0\,M$_\odot$, 7.5\,M$_\odot$ Z=0.001 models and 4, 25 for the 7.0\,M$_\odot$, 7.5\,M$_\odot$ Z=0.0001 models, respectively. Although we expect C2DU events at all metallicities, they have a stronger signature at lower metallicity, as they enrich the surface with similar amounts of material, but into an initially lower metallicity environment.
Fig.~\ref{fig-surf} also shows the large helium contribution from 2DU, with enhancement from the initial mass fraction of 0.248 to $\sim$ 0.35.

In comparison to a standard 2DU event which depletes \chem{12}C,\chem{16}O and \chem{22}Ne, corrosive 2DUs result in large surface enrichments of \chem{12}C, and in some cases \chem{16}O, \chem{18}O and \chem{22}Ne. If the model is very metal poor this dredge-up of \chem{12}C may lead to the C/O ratio exceeding unity and therefore to the formation of a carbon star. In both standard and corrosive 2DU events \chem{23}Na is significantly enhanced. 

The initial and 2DU/C2DU/CO2DU surface abundances (in mass fraction) can be found in Table~\ref{initial2du} in Appendix A.

\subsection{Thermally Pulsing Super-AGB phase}\label{sec-tp}

During the thermally pulsing phase the surface composition is altered by both HBB and 3DU. Table~\ref{table1} highlights important quantities for this phase of evolution.

\begin{figure*}
\begin{center}
\includegraphics[width=13.5cm,angle=0]{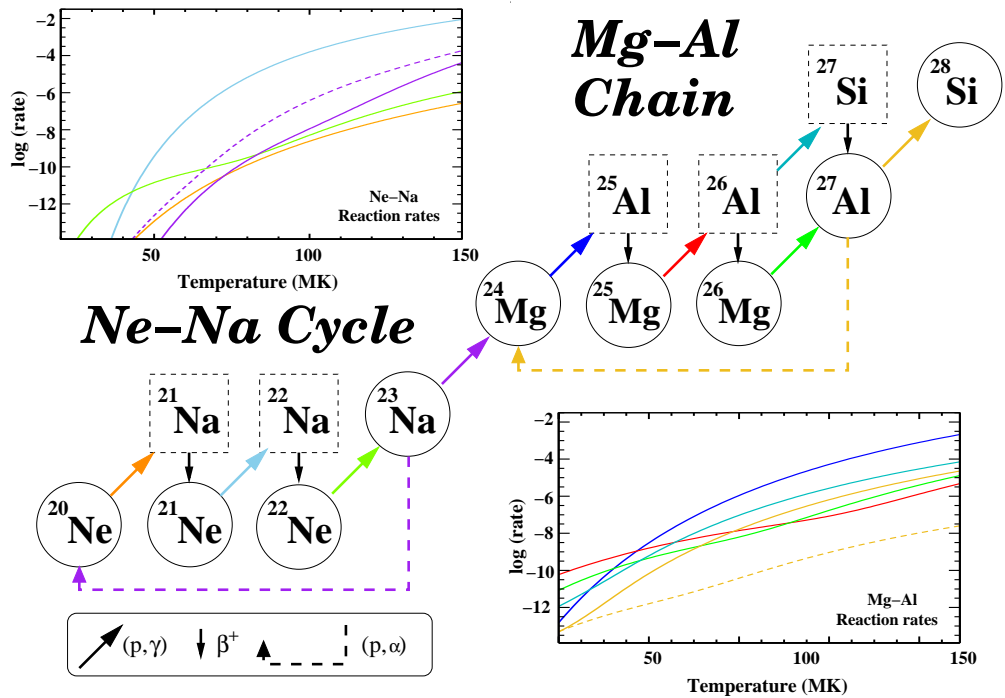}
\caption{Schematic (with atomic mass number on the x-axis and proton number on the y-axis) for the Ne-Na cycle and Mg-Al chains with reaction rates in the temperature range 30-150MK. Solid circles denote stable isotopes whilst dashed squares show unstable isotopes. Reactions indicated by a coloured arrow have their rate plotted in the same colour. The terrestrial half-life of radioactive decays are \chem{21}Na $\sim$ 22.5sec, \chem{22}Na $\sim$ 2.60yrs, \chem{25}Al $\sim$ 7.25sec, \chem{26}Al $\sim$ 0.717 Myrs and \chem{27}Si $\sim$4.16sec. 
The reaction rates used in the Ne-Na cycle for this study are: \chem{20}Ne(p,$\gamma$)\chem{21}Na from \texttt{NACRE}, \chem{21}Ne(p,$\gamma$)\chem{22}Na from \protect\cite{ili01}, \chem{22}Ne(p,$\gamma$)\chem{23}Na from \protect\cite{hal02}, \chem{23}Na(p,$\gamma$)\chem{24}Mg and \chem{23}Na(p,$\alpha$)\chem{20}Ne from \protect\cite{hal04}, while all the reaction rates in the Mg-Al chain are from \protect\cite{ili01}.}
\label{fig-nena} \end{center}
\end{figure*}

The very high temperatures achieved at the base of the convective envelope result in HBB with activation of the CNO, Ne-Na, Mg-Al and Si proton capture reaction pathways (Fig.~\ref{fig-nena}).
Generally, models of lower metallicity and/or higher mass, i.e. those having at a given mass a more massive core, attain higher temperatures at the base of the convective envelope and undergo more advanced nucleosynthesis. 
With our preferred mass-loss rate \citep{vas93} we typically find that lower metallicity super-AGB stars of a given initial mass have a longer super-AGB phase and undergo more thermal pulses with associated 3DU events than their higher metallicity counterparts (see Table~\ref{table1}). 

The nucleosynthesis that takes place within a thermal pulse convective zone consists of the creation of \chem{12}C, \chem{16}O, \chem{22}Ne, \chem{25}Mg, \chem{26}Mg and to a lesser extent \chem{24}Mg, \chem{20}Ne, \chem{23}Na and \chem{21}Ne. This material is subsequently dredged to the surface via 3DU. We find third dredge-up efficiency\footnote{$\lambda$ = $\Delta$ $M_{\rm{dredge}}$/$\Delta$ $M_{\rm{H}}$, where $\Delta$ $M_{\rm{H}}$ is the increase in the core mass during the interpulse phase and $\Delta$ $M_{\rm{dredge}}$ is the mass of dredged-up material.} for this mass and metallicity range is quite high with $\lambda$$\sim$0.7-0.8. The total mass dredged-up due to 3DU ($M_{\rm{Dredge}}^{\rm{Tot}}$) in any of these super-AGB stars is at most $\sim$ 0.3\,M$_\odot$ (of which $\approx$ 30 per cent by mass is \chem{12}C and $\approx$ 2 per cent \chem{16}O), with an approximate halving of dredged up material for every 0.5\,M$_\odot$ increase in initial mass (Table~\ref{table1}). Although this is quite a small amount, at these low metallicities the third dredge-up can lead to an increase of over an order of magnitude in the surface abundance of metal-rich material. It must be noted when considering the importance of 3DU to the overall stellar yield that earlier the C2DU also contributed to a significant envelope enrichment in metal.

As illustrated in Fig.~\ref{fig-surf}, the surface abundance of helium increases along the TP-(S)AGB phase from a combination of HBB and 3DU, with the majority of this increase, about 70 per cent, due to HBB.  
The stars become lithium rich for a short time ($\approx$ 20,000 years) at the start, or just prior to the TP-(S)AGB phase before Li is efficiently destroyed once the \chem{3}He supply has been fully depleted. 

The behaviour of the C, N and O isotopes is dictated mainly by CNO cycling during HBB. The \chem{12}C is initially depleted and converted into \chem{13}C then \chem{14}N. The addition of fresh \chem{12}C from 3DU episodes results in increasingly higher \chem{14}N abundances due to this CNO cycling. Both nitrogen isotopes \chem{14}N and \chem{15}N increase steadily whilst all three oxygen isotopes \chem{16}O, \chem{17}O, \chem{18}O  see substantial destruction during the TP-(S)AGB phase. The \chem{12}C/\chem{13}C ratio reaches its equilibrium value as do \chem{17}O/\chem{16}O and \chem{15}N/\chem{14}N.

Very hot HBB via the Ne-Na cycle (Fig.~\ref{fig-nena}) will result in destruction of \chem{21}Ne, \chem{22}Ne and \chem{23}Na to the benefit of \chem{20}Ne while the 3DU contributes to the envelope enrichment of \chem{22}Ne and to a lower extent \chem{23}Na and \chem{20}Ne. 

If temperatures at the base of the convective envelope are sufficiently high ($T_{\rm{BCE}}$ $>$ 80MK), which is the case in all our computed models, the Mg-Al chain (Fig.~\ref{fig-nena}) is activated. The \chem{24}Mg(p,$\gamma$)\chem{25}Al($\beta^{+}$)\chem{25}Mg reaction proceeds rapidly and converts the majority of the initially most abundant magnesium isotope \chem{24}Mg to \chem{25}Mg. The \chem{25}Mg subsequently depletes via \chem{25}Mg(p,$\gamma$)\chem{26}Al(p,$\gamma$)\chem{27}Si($\beta^{+}$)\chem{27}Al with a large increase in \chem{27}Al.
All three magnesium isotopes are synthesized within the helium burning intershell region and are brought back into the envelope by 3DU events. The \chem{24}Mg is created primarily via the \chem{20}Ne($\alpha$,$\gamma$) channel, and only when the envelope abundance has been sufficiently depleted via HBB that this small production will be seen as a noticeable increase at the surface (Fig.~\ref{fig-surf}). The \chem{26}Mg is created via both the \chem{22}Ne($\alpha$,$\gamma$) and \chem{25}Mg(n,$\gamma$) channels. The silicon isotope \chem{28}Si shows a small increase in the later phases of evolution from \chem{27}Al(p,$\gamma$)\chem{28}Si (see also \citealt{sie08b}).

Third dredge-up events also increase the surface abundance of our s-process proxy isotope $g$, with the longer lived and/or lower-mass models being enhanced by a factor of over 100 from their initial value. 

 \begin{table}\setlength{\tabcolsep}{1.0pt}
 \caption{Selected model characteristics where: $M_{\rm{ini}}$ is the initial mass, $T_{\rm{BCE}}^{\rm{Max}}$ is the maximum temperature at the base of the convective envelope; $M_{\rm{Dredge}}^{\rm{Tot}}$ is the total mass of material dredged to the surface due to 3DU; $M_{\rm{2DU}}$ is the post 2DU core mass; $M_{\rm{C}}^{\rm{F}}$ is the final computed core mass; $M_{\rm{env}}^{\rm{F}}$ is the final envelope mass, $\langle\dot{M}\rangle$ is the average mass-loss rate during the TP-(S)AGB phase; $N_{\rm{TP}}$ is the number of thermal pulses and $\tau_{\rm{(S)AGB}}$ is the duration of the thermally pulsing (S)AGB phase. The B indicates models produced using the B95 mass-loss rate, $\alpha$ indicates the model with an increased mixing-length $\alpha$ value of 2.1, whilst models indicated with O are those using variable composition low-T molecular opacities. Note that $n(m)= n \times 10^{m}$.}\label{table1}
 \begin{center} \begin{tabular}{lccccccrc} \hline
$M_{\rm{ini}}$&$T_{\rm{BCE}}^{\rm{Max}}$&$M_{\rm{Dredge}}^{\rm{Tot}}$ &$M_{\rm{2DU}}$ & $M_{\rm{C}}^{\rm{F}}$&$M_{\rm{env}}^{\rm{F}}$&$\langle\dot{M}\rangle$ &$N_{\rm{TP}}$&$\tau_{\rm{(S)AGB}}$  \\
(M$_\odot$)& (MK) & (M$_\odot$)&  (M$_\odot$)& (M$_\odot$)  &  (M$_\odot$) &(M$_\odot$/yr)&& (yrs) \\
\hline
 \multicolumn{9}{c}{Z=0.001}   \\
 \hline 
6.5          &116 &8.32(-2) &1.05 &1.07 &0.58&2.84(-5)&118&1.69(5)\\
6.5B         &108 &2.01(-2) &1.05 &1.06 &0.62&9.82(-5)&38 &4.78(4)\\ 
7.0          &120 &3.97(-2) &1.13 &1.14 &0.56&6.67(-5)&126&7.83(4)\\
7.0B         &118 &1.07(-2) &1.13 &1.14 &0.38&2.11(-4)&47 &2.45(4)\\
7.0B$\alpha$ &119 &4.37(-3) &1.13 &1.14 &0.09&4.17(-4)&28 &1.23(4)\\ 
7.5          &122 &2.64(-2) &1.20 &1.21 &0.73&1.03(-4)&171&5.22(4)\\
7.5B         &120 &5.81(-3) &1.20 &1.21 &0.47&3.56(-4)&56 &1.43(4)\\ 
\hline \multicolumn{9}{c}{Z=0.0001} \\  \hline
6.5          &120 &2.72(-1) &1.06 &1.10 &0.24&1.02(-5)&439&5.00(5)\\  
6.5O         &119 &1.28(-1) &1.06 &1.08 &0.77&1.81(-5)&214&2.55(5)\\
6.5B         &119 &1.46(-2) &1.06 &1.06 &0.24&1.39(-4)&35 &3.68(4)\\
7.0          &128 &1.23(-1) &1.14 &1.17 &0.47&2.37(-5)&459&2.24(5)\\
7.0O         &129 &6.30(-2) &1.14 &1.17 &0.76&4.00(-5)&269&1.26(5)\\
7.0B         &126 &5.88(-3) &1.14 &1.15 &0.19&3.29(-4)&38 &1.58(4)\\
7.5          &127 &5.41(-2) &1.21 &1.23 &0.43&5.48(-5)&445&1.06(5)\\
7.5O         &127 &3.03(-2) &1.21 &1.22 &0.70&8.83(-5)&248&6.23(4)\\
7.5B         &122 &2.30(-3) &1.21 &1.21 &0.47&6.93(-4)&35&6.82(3)\\ 
\hline \end{tabular} \end{center} \end{table}

\section{Yields}\label{yields}

We calculate nucleosynthetic (net) yields (in M$_\odot$) using the following expression

\begin{equation}
 M_{i} = \int_{0}^{\tau} \left[ X(i) - X_{\rm{ini}}(i)\right] 
\dot{M}(t) dt,
\label{yield}
\end{equation}

where $X(i)$ and $X_{\rm{ini}}(i)$ are the current and initial mass fractions of species $i$ respectively, $\tau$ is the stellar lifetime and $\dot{M}(t)$ is the mass-loss rate.

Another useful nucleosynthetic quantity is $\langle X(i)\rangle$, the average mass fraction of species $i$ expelled into the wind, calculated as $\langle X(i)\rangle$ = $M_{i}$/$\Delta$M + $X_{\rm{ini}}(i)$, where $\Delta$M is the total mass ejected in the stellar wind.

We have produced two sets of nucleosynthetic yields with differing mass-loss rates during the super-AGB phase. In our standard set we use the VW93 rate and in the second set we use the more rapid mass-loss rate of \cite{blo95} with $\eta$=0.02 (see Section~\ref{sec-ml} for details). We have also calculated additional yields for three test cases to explore the uncertainties associated with (1) an increased mixing-length parameter $\alpha_{mlt}$, (2) the inclusion of the low temperature variable composition molecular opacities from \cite{led09} and (3) the use of updated nuclear reaction rates. The analysis of these simulations is presented in Section~\ref{sec-uncertainties}.

\begin{figure*}
\resizebox{\hsize}{!}{\includegraphics{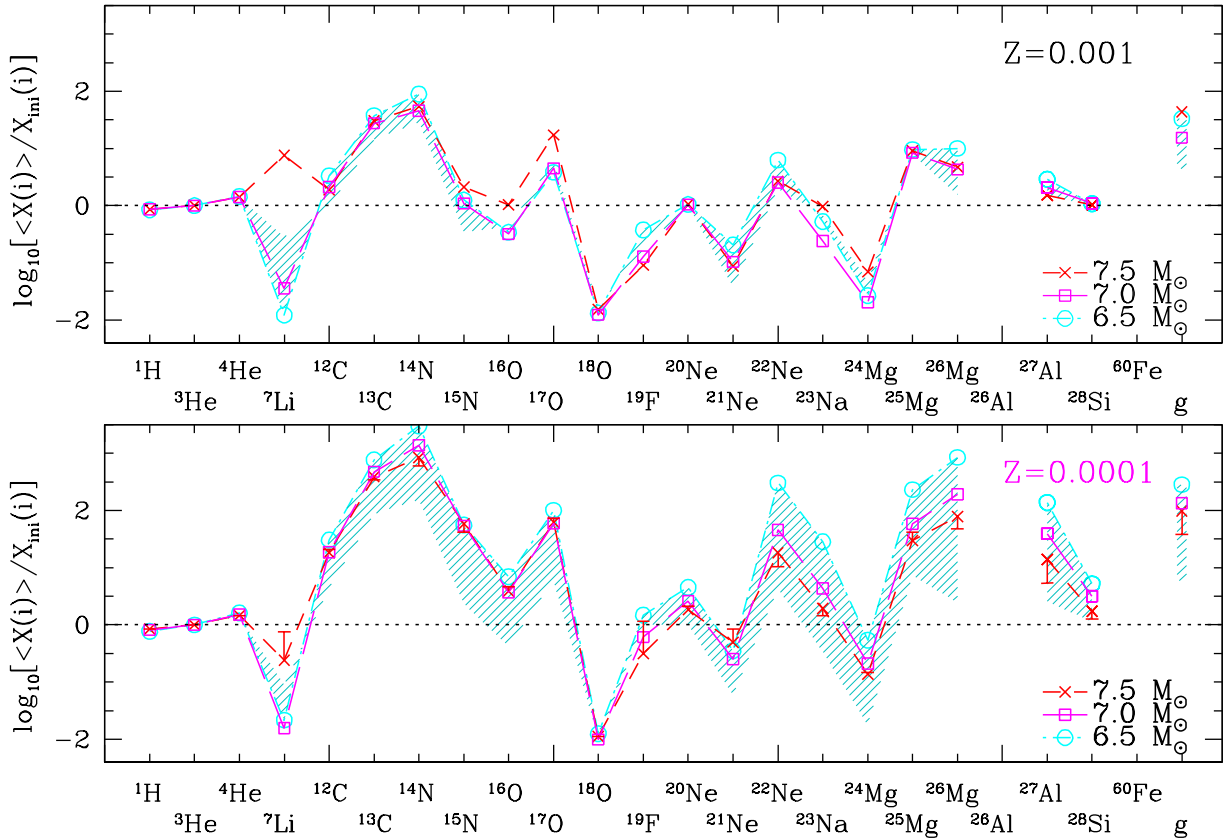}}
\caption{Production Factor log$_{10}$[$\langle$X(i)$\rangle$/X$_{\rm{ini}}$(i)] of selected species. $g$ is used as a proxy for elements heavier than iron. The models presented in this figure with symbols are our standard VW93 mass-loss rate set. The extent of the cyan shaded regions represents the yields from the B95 mass-loss rate 6.5\,M$_\odot$ models. The error bars on the 7.5\,M$_\odot$ Z=0.0001 symbols show the production factor for the variable composition molecular opacity test case (most are too small to be seen).}
\label{fig-yield}
\end{figure*}

The nucleosynthetic results from all these models are presented as isotopic production factors $\log_{10}[\langle X(i)  \rangle/X_{\mathrm{ini}}(i)]$ in Fig.~\ref{fig-yield}, and as elemental yields, defined as the average composition of the ejecta $\langle X(i)\rangle$ expressed via [X/Fe]\footnote{where [A/B]=log$_{10}$($n$(A)/$n$(B))$_*$-log$_{10}$($n$(A)/$n$(B))$_\odot$} in Table~\ref{table2}, with the latter quantity allowing for more direct comparison to observations.

Near the end of the evolution, instabilities in the convective envelope \citep{woo86,lau12} prevent convergence and halt calculations. Because this phase of evolution is uncertain, and with the non-negligible amount of envelope mass left at that time (see Table~\ref{table1}), we provide two subsets of yields for the models.\footnote{In Table~\ref{tableappend2} of Appendix A we give a sample format of the yield tables.} In the first subset of yields (our standard) we assume the remaining envelope is expelled with the same composition as that of the last computed model, whilst in the second subset of yields we include extrapolated thermal pulses (as described in Paper II in our series) to account for possible nucleosynthesis in the case that evolution had continued past the convergence difficulties. We discuss these differences in Section~\ref{subsec-extrap}.

We find many similarities with our more metal rich super-AGB star yields (Paper II), so here we mainly highlight the features which are pertinent and/or only occur at the low metallicities studied here.
 
Helium is produced in large quantities, with the bulk coming from 2DU, and further increases primarily from efficient HBB, although there is some 3DU contribution.
The yields of \chem{7}Li are strongly dependent on the rate of mass loss, with only the 7.5\,M$_\odot$ Z=0.001 model producing a positive yield. 
In contrast to the more metal rich models, all of our standard low metallicity models have positive \chem{12}C yield either due to a larger number of 3DU events or because of the stronger impact of corrosive 2DU events. There are large enhancements of \chem{13}C and \chem{14}N due to HBB. With high $T_{\rm{BCE}}$ the \chem{15}N created via the CN cycle reaches equilibrium with \chem{14}N and results in a positive yield. 

The elemental yield of oxygen is negative for the 6.5\,M$_\odot$ and 7.0\,M$_\odot$ Z=0.001 models, with destruction of all three main oxygen isotopes \chem{16}O, \chem{17}O and \chem{18}O from HBB.
 Conversely, at Z=0.0001 the 6.5\,M$_\odot$ and 7.0\,M$_\odot$ models have positive elemental oxygen yields due to the surface enrichment of \chem{16}O from 3DU events. Although the intershell is composed of only 1-3 percent \chem{16}O, due to the contrast with the very low initial abundance, this enrichment is enough to overpower the effect of HBB. The contribution from the CO2DU results in a positive yield for the 7.5\,M$_\odot$ models of both metallicities.

In such a hot environment \chem{19}F is efficiently destroyed, however, in some models the efficient 3DU at the end of the evolution after cessation of HBB can lead to slight positive yields. The elemental Ne yield is positive from the increase of \chem{20}Ne and \chem{22}Ne although \chem{21}Ne is effectively destroyed via proton capture in HBB. The elemental Mg yields are all positive due to the increase of \chem{25}Mg from efficient HBB and 3DU and \chem{26}Mg almost solely from 3DU. A minor contribution to the \chem{26}Mg yield comes from the decay of \chem{26}Al.

The radioactive isotope \chem{26}Al is created in large quantities $\sim$ 0.1--3$\times$10$^{-5}$\,M$_\odot$, similar to the values found by \cite{sie08b}. With the high $T_{\rm{BCE}}$ and long duration of the TP-(S)AGB phase, the \chem{26}Al(p,$\gamma$)\chem{27}Si($\beta^{+}$)\chem{27}Al channel opens which bypasses the \chem{26}Mg. 
Aluminium yields are always positive, with the main contribution to \chem{27}Al from HBB nucleosynthesis.

Silicon is produced in all models from a combination of both 3DU and HBB.  The contribution from 3DU to \chem{28}Si production drops from about 80 per cent in the lowest mass model to about 20 per cent in the most massive model. Whilst \chem{28}Si production is only marginal at Z=0.001, it increases with decreasing metallicity, which results in large enhancements at Z=0.0001. The amount of heavier silicon isotopes, \chem{29}Si and \chem{30}Si, is also greatly increased due to proton captures, and via neutron captures in the intershell region during a thermal pulse. 
Although there is a substantial production of Si in our models, the overall contribution to the interstellar medium, compared to that from more massive star (supernova) yields is quite small.\footnote{The massive star contribution to interstellar abundance of Si for these low metallicity models is more than 4 orders of magnitude greater than from super-AGB stars, assuming the yields are weighted via a standard Kroupa IMF \citep{kro93}, and using the massive star yields from \cite{kob06}.} The production of alpha-elements O, Ne and Mg is also relatively minimal in super-AGB stars of these metallicities compared to their higher mass counterparts.

The main production channel of \chem{31}P is via the reaction \chem{27}Al($\alpha$,$\gamma$) within the thermal pulse, with this material subsequently mixed to the surface through 3DU events. Phosphorus production, is minimal at Z=0.001, similar to the silicon isotopes, but becomes quite large in the lower mass Z=0.0001 models, with increases by up to 1.7 dex in the 6.5\,M$_\odot$ model.

Our s-process proxy species $g$ increases with decreasing metallicity and shows up to 2 dex increase in the Z=0.0001 models. We explore this s-process production in super-AGB stars in a forthcoming work. 

\begin{table*}\setlength{\tabcolsep}{3.5pt}
\begin{center} 
\caption{Nucleosynthetic quantities: the average mass fraction of helium in the wind $\langle$Y$\rangle$, $\langle$log$_{10}$$\epsilon$(\chem{7}Li)$\rangle$ where log$\epsilon$(\chem{7}Li) = log (n[Li]/n[H]) + 12 and the elemental yields via [X/Fe]. The B in the initial mass column indicates models run with the B95 mass-loss rate with $\eta$=0.02, whilst O refers to the inclusion of low temperature variable composition molecular opacities. $R_{\rm{CNO}}$ is the ratio of the average number fraction of C+N+O ejected in the wind to the initial C+N+O. Initial helium mass fractions are Y= 0.2492 and 0.248 for Z=0.001 and 0.0001 respectively. Initial log$_{10}$$\epsilon$(\chem{7}Li) is set at the Spite plateau's value $\sim$ 2.36. $\alpha$ represents modified mixing-length value from 1.75 to 2.1. RR refers to use of the reaction rate network JINA V2.0.}
\label{table2}
\begin{tabular}{lcccccccccccccc}
\hline
$M_{\rm{ini}}$&$\langle$Y$\rangle$&$\langle$log$_{10}$$\epsilon$(\chem{7}Li)$\rangle$&[C/Fe]&[N/Fe]&[O/Fe]&$R_{\rm{CNO}}$&[F/Fe]&[Ne/Fe]&[Na/Fe]&[Mg/Fe]&[Al/Fe]&[Si/Fe]&[$g$/Fe]&\\ 
\hline 
\multicolumn{14}{c}{Z=0.001}   \\ \hline
6.5        & 0.359 &0.46& 0.60&1.98&-0.43&8.32  &-0.42&0.15&-0.27&0.34 &0.47& 0.04&1.53\\
6.5B       & 0.350 &1.81& 0.05&1.49&-0.34&2.87  &-0.76&0.03&-0.16&0.02 &0.17& 0.01&0.69\\
7.0          & 0.354 &0.92& 0.41&1.69&-0.47&4.49  &-0.89&0.06&-0.62&0.14 &0.32& 0.04&1.19\\ 
7.0RR        & 0.354 &0.99& 0.41&1.69&-0.48&4.54  &-1.56&0.06&-0.43&0.12 &0.44& 0.03&1.41\\
7.0B         & 0.350 &2.45&-0.01&1.31&-0.31&2.07  &-1.07&0.02&-0.24&0    &0.17& 0.01&0.53\\
7.0B$\alpha$ & 0.350 &2.36&-0.07&1.23&-0.30&1.78  &-1.03&0.02&-0.19&-0.01&0.16& 0.01&0.41\\
7.5        & 0.354 &3.25& 0.37&1.76& 0.05&5.53  &-1.04&0.07&-0.01&0.18 &0.18& 0.02&1.64\\
7.5B       & 0.353 &3.70& 0.21&1.55& 0.10&3.77  &-0.78&0.03& 0.35&0.02 &0.11& 0.01&0.71\\
\hline
\multicolumn{14}{c}{Z=0.0001}  
 \\ \hline
6.5        & 0.402 &0.75& 1.63&3.52& 0.89&249.52&0.19 &1.41& 1.48& 2.09&2.16& 0.82&2.47\\
6.5O       & 0.370 &0.37& 1.58&3.15& 0.55&114.74&0.35 &0.89& 0.89& 1.40&1.37& 0.33&2.05\\
6.5B       & 0.351 &1.31& 0.75&2.22&-0.34&13.99 &-0.67&0.10&-0.48& 0.03&0.41& 0.05&0.77\\
7.0          & 0.377 &0.58& 1.41&3.17& 0.61&116.82&-0.21&0.75& 0.65& 1.44&1.60& 0.52&2.13\\
7.0O         & 0.362 &0.60& 1.37&2.87& 0.32&61.17 &-0.44&0.44& 0.29& 0.94&0.98& 0.30&1.78\\
7.0B         & 0.351 &2.11& 0.64&1.97&-0.39&8.36  &-0.93&0.04&-0.47&-0.01&0.26& 0.05&0.54\\
7.5        & 0.363 &1.76& 1.37&2.95& 0.62&74.5  &-0.50&0.47&0.29 &1.08 &1.14& 0.25&1.99\\
7.5O       & 0.356 &2.26& 1.41&2.81& 0.70&57.72 & 0.06&0.33&0.16 &0.88 &0.74& 0.10&1.58\\
7.5B       & 0.352 &3.68& 0.99&2.46& 0.89&28.43 &-0.76&0.04&0.38 &0.04 &0.08& 0.01&0.42\\
\hline
\end{tabular}
\end{center}
\end{table*}

\subsection{Uncertainties}\label{sec-uncertainties}

Super-AGB star modelling is subject to a large range of uncertainties. We have performed a series of tests to gauge the sensitivity of our yield results to various physical assumptions.

\subsubsection{Mass-loss rate}\label{sec-ml}

The mass-loss rate for low-metallicity massive AGB stars has always been highly contentious and is generally considered to be a decreasing function of metallicity. However, a growing collection of work, both theoretical \cite[e.g.][]{mat08b,wac08} and observational \cite[e.g.][]{gro09,lag08} suggest this may not always be the case, due to factors such as differences in dust production, higher luminosities and the importance of carbon in low metallicity stellar environments.
Another relevant point when considering mass-loss is the surface composition during the AGB phase. Corrosive 2DU, dredge-out or 3DU events in lower metallicity models can considerably enrich the surface composition to metallicities comparable to/or even greater than solar composition \citep{gil13}. We use these as our justifications for using ``normal'' mass-loss rates in this work. 

Our standard mass-loss rate is an empirical prescription derived by \cite{vas93} from CO microwave observations of both carbon-rich and oxygen-rich AGB stars in the Galaxy and Large Magellanic Cloud. 
We have also chosen to test the effect of using the \cite{blo95} mass-loss rate. This rate is popular amongst low-metallicity AGB star modellers \cite[e.g.][]{her04a,ven13} and is based on the prescription of \cite{rei75}, modified to take into account the 2D hydrodynamic models of low-mass Mira variables \citep{bow88}. The free parameter in the B95 rates is calibrated to various observable quantities, and we use the value $\eta$=0.02 derived by \cite{ven00} based on Li-rich giants in the Large Magellanic Cloud.\footnote{This $\eta$ calibration was derived using models with the more efficient full spectrum turbulence convective approach, meaning a larger $\eta$ value would be required in our models to match these Li observations. However, at these lower metallicities the choice of $\eta$ is unconstrained.}

Table~\ref{table1} highlights the greatly reduced duration of the TP-(S)AGB phase when the B95 mass-loss rate is used compared to the VW93 mass-loss rate. 

These mass-loss rates diverge with decreasing metallicity with over a factor of 10 difference in the duration of the TP-(S)AGB phase for the Z=0.0001 models.
 The B95 models of metallicity Z=0.001 experience a factor of three times fewer thermal pulses, whilst the Z=0.0001 models have more than 10 times fewer thermal pulses. This greater difference is related to the different dependencies on stellar parameters within the mass-loss prescriptions; with the B95 having a large luminosity exponent, whilst the VW93 rate is more radius dependent. We note here, that low metallicity stars are more compact and have a higher luminosity.

In Fig.~\ref{fig-surf} (upper four panels) the B95 mass-loss rate results are plotted over the standard VW93 case for the 7.0\,M$_\odot$ Z=0.001 model. The surface abundance evolution of both models is strikingly similar for the majority of the evolution. The more prominent upturn of \chem{12}C near the end of evolution in the B95 case simply reflects a dilution effect of the 3DU in the smaller remaining envelope mass.  

\begin{figure}
\resizebox{\hsize}{!}{\includegraphics{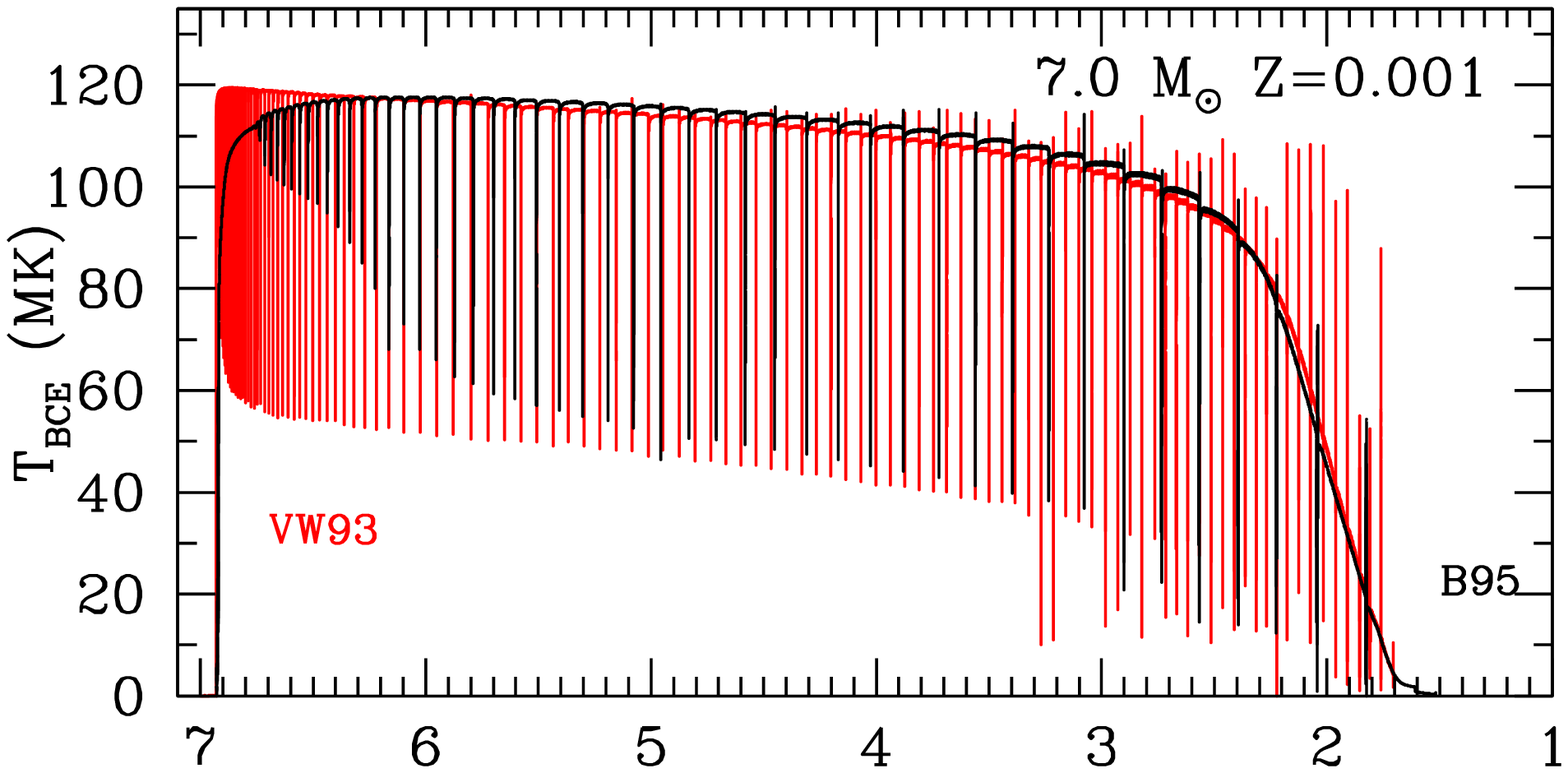}}
\resizebox{\hsize}{!}{\includegraphics{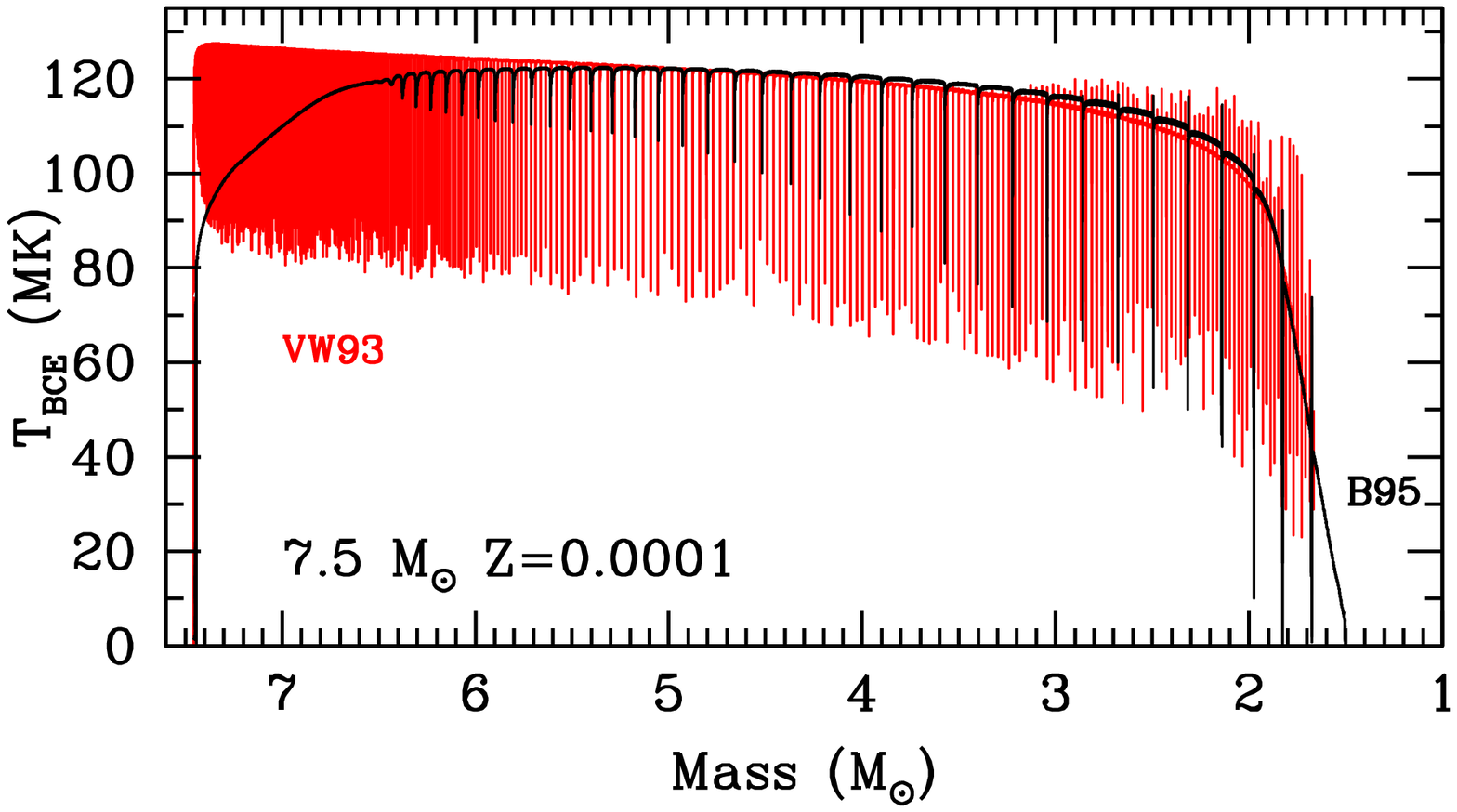}}
\caption{Temperature at the base of the convective envelope as a function of total stellar mass for the 7.0\,M$_\odot$ Z=0.001 (top panel) and 7.5\,M$_\odot$ Z=0.0001 (bottom panel) models with VW93 (red) and B95 (black) mass-loss rates. }
\label{fig-tbce}
\end{figure}

In Fig.~\ref{fig-tbce} we plot the temperature at the base of the convective envelope as a function of total mass for the VW93 and B95 mass-loss rate models for the 7.0\,M$_\odot$ Z=0.001 (top panel) and 7.5\,M$_\odot$ Z=0.0001 (bottom panel) models. In this last panel when the B95 prescription is used the stellar wind had already stripped over 1\,M$_\odot$ of envelope prior to the first thermal pulses. This highlights the very rapid $\approx$ 7 $\times 10^{-4}$ M$_\odot$ yr$^{-1}$ mass-loss rate, which results in an extremely short TP-(S)AGB phase of just under 7000 years (Table~\ref{table1}).  
Apart from the early AGB phase the subsequent evolution of $T_{\rm{BCE}}$ for models with different mass-loss rates shows striking similarities. 

The largest variation in yields is found in models most affected by 3DU enrichment, these being the least massive and/or more metal poor models where the amount of dredged material $M_{\rm{Dredge}}^{\rm{Tot}}$ is greater (Table~\ref{table1}). In Fig.~\ref{fig-yield} the extent of the shaded regions represents the yields from the B95 mass-loss rate 6.5\,M$_\odot$ models for both Z=0.001 and 0.0001. Clearly seen is the larger discrepancy in yield prediction results with differing mass-loss rates as the metallicity decreases. 
Due to the reduced duration of the TP-(S)AGB phase with the B95 prescription, fewer 3DU episodes take place and less time is available to nuclearly process the envelope via HBB. The impact of this truncated thermally pulsing phase mostly affects the surface abundances of \chem{22}Ne, \chem{23}Na, \chem{24,25,26}Mg and \chem{27}Al, with an $\approx$ 2 orders of magnitude decrease in their yields. In the Z=0.0001 models, variations of over 1 order of magnitude in the yields results of \chem{14}N, \chem{16}O and the s-process proxy isotope $g$ between mass-loss rates are found. Although lithium yields are highly dependent on the mass-loss rate, even with the rapid B95 mass-loss prescription, the 6.5\,M$_\odot$ models still have negative \chem{7}Li yields. We also note a substantial reduction of the helium yield for the 6.5\,M$_\odot$ Z=0.0001 model when the B95 rate is used from $\langle$Y$\rangle$ = 0.402 to 0.351.  

\subsubsection{Mixing-length parameter}

The thermodynamic conditions at the base of the convective envelope are affected by the choice of convective mixing approach. In this work we use the mixing-length theory (MLT) with a solar calibrated $\alpha_{\rm{mlt}}$ value of 1.75. Some studies have used other convective theories in AGB models such as full spectrum turbulence (FST; \citealt{can91}), or modified mixing-length theory \citep{woo11a}. These prescriptions correspond to more efficient mixing (larger values of $\alpha_{\rm{mlt}}$) than in the standard MLT case. Although simply increasing $\alpha_{\rm{mlt}}$ is not a direct analogue for implementing these more sophisticated (albeit not necessarily more accurate) convective approaches, it can however, mimic the trends/results from these works. To test the sensitivity of our results to convective modelling, the 7.0\,M$_\odot$ Z=0.001 model with B95 mass-loss rate was recomputed with an increased $\alpha_{mlt}$ value of 2.1. This combination of increased mass-loss and $\alpha_{mlt}$ was chosen for its similarity to the works by \cite{ven05} and \cite{ven13}. An increase in $\alpha_{\rm{mlt}}$ leads to higher luminosities and higher temperatures at the base of the convective envelope. 
When $\alpha_{\rm{mlt}}$ is increased and combined with the high mass-loss rate the duration of the TP-(S)AGB phase is halved (Table~\ref{table1}). The nucleosynthesis yields however, are only slightly altered, mainly with a small decrease in N, $R_{\rm{CNO}}$ (the ratio of the average number fraction of C+N+O ejected in the wind to the initial C+N+O) and $g$ (Table~\ref{table2}). The cause of this similarity in results is twofold. First, due to the rapid unifying of the surface composition due to HBB once the star reaches the TP-(S)AGB phase the slight difference in $T_{\rm{BCE}}$ does not greatly affect the nucleosynthesis. Moreover, at this large core mass, the contribution to surface enrichment from each individual third dredge-up event is very modest, so the change in number of pulses makes little difference. 

\subsubsection{Nuclear reaction rates}

Another major source of uncertainty in super-AGB star nucleosynthesis modelling is the nuclear reaction rates, with the most important reactions in this regard being: the Ne-Na cycle and Mg-Al chain during HBB \citep[e.g.][]{arn99,izz07,van08,sie08b} and the \chem{22}Ne+$\alpha$ reactions within the thermal pulse convective zone \citep{kar06a,lon12}. 

The nucleosynthesis for the 7.0\,M$_\odot$ Z=0.001 model was recomputed in \textsc{monsoon} to test the effects of using the updated reaction rate network from JINA V2.0 compared to V1.0 \citep{jin10}. The main modifications concern the Ne-Na, Mg-Al proton capture rates updated to the \cite{ili10} set, as well as the \chem{22}Ne($\alpha$,n)\chem{25}Mg and \chem{22}Ne($\alpha$,$\gamma$)\chem{26}Mg from \cite{kar06a} also updated to \cite{ili10}. The yields of the main elements (e.g. He, Li, C, N, O) show remarkable agreement between the updated and standard results (Table~\ref{table2}). The differences are isolated to a small selection of isotopes; with $+0.2$ dex in $g$, \chem{23}Na and \chem{27}Al, $-0.1$ dex in \chem{26}Mg, with the largest change being $-0.7$ dex in \chem{19}F. These changes are expected, with the faster \chem{22}Ne($\alpha$,n)\chem{25}Mg reaction rate from Iliadis leading to larger production of the neutron-rich isotopes $g$ and \chem{23}Na. The updated \chem{26}Mg(p,$\gamma$)\chem{27}Al rate is also faster and has resulted in greater production of \chem{27}Al to the deficit of \chem{26}Mg. Finally, the reduction in \chem{19}F yield in this model is due to the more rapid destruction channel \chem{19}F(p,$\gamma$)\chem{20}Ne from \cite{cf88} compared to \texttt{NACRE}. We would expect similar (small) changes to the yields across this mass and metallicity range. 

\subsubsection{Variable low temperature opacities}

The use of molecular opacities allowing for variations in the C and N content at low temperature are crucial to correctly model stellar envelopes, especially when the carbon to oxygen number ratio exceeds unity \cite[e.g.][]{mar02,ven10b}. When the star becomes C-rich, and the effective temperature is sufficiently low ($T_{\rm{eff}}$ $\la$ 4000 K), the molecular chemistry changes dramatically, with formation of carbon bearing molecules producing a substantial increase in the opacity. This leads to an expansion of the stellar envelope and enhanced mass-loss rate resulting in a shorter TP-(S)AGB phase. 

Since the majority of super-AGB stars of near solar or moderately low metallicity either spend the entirety of their lives oxygen rich due to HBB or only become carbon rich due to 3DU late in their evolution when already in the superwind phase, one may expect that the inclusion of CN variable opacities would have minimal effect. This is not the case however, for some very metal-poor super-AGB stars which can become carbon stars either by undergoing a C2DU event, repeated 3DU enrichment or from HBB very efficiently depleting oxygen.

\begin{figure}
 \resizebox{\hsize}{!}{\includegraphics{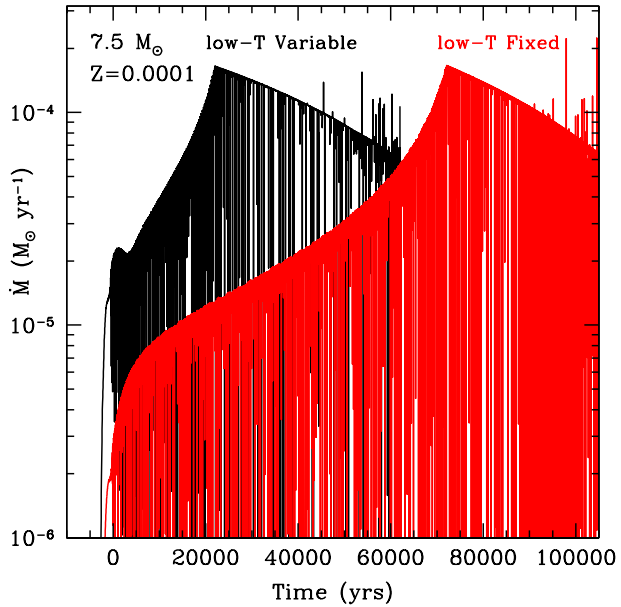}}
 \caption{Mass-loss rate as a function of time for the 7.5\,M$_\odot$ Z=0.0001 with low-T fixed (red) and low-T variable (black) composition molecular opacities.}
 \label{fig-opacs}
 \end{figure}

Our representative 7.5\,M$_\odot$ Z=0.0001 VW93 model was recomputed using the variable C, N low temperature opacities of \cite{led09}.
The resultant evolutionary and nucleosynthetic information is provided in Tables~\ref{table1} and \ref{table2} with these models denoted by an O. As with previous studies focusing on low-mass AGB stars \cite[e.g.][]{cri09,lug12a}, the duration of the TP-SAGB phase $\tau_{SAGB}$ is significantly reduced, in this case by approximately 50 per cent. The evolution of the surface abundances and mass-loss rate are shown in Fig.~\ref{fig-surf} (lower four panels) and Fig.~\ref{fig-opacs} respectively. The initial composition is scaled solar, whilst it is known that stars of such metallicities generally have $\alpha$ (mostly oxygen) enhancements. If this effect had been taken into account, the C/O ratio may have remained below one and the effects of varying opacity reduced. 
The early peak in $\dot{M}$ at t $\simeq$ 0 in the low-T variable composition opacity test model corresponds to the first occurrence of C/O $>$ 1. Later, although the C/O ratio dips below unity (Fig.~\ref{fig-surf}), the mass-loss rate has already entered the superwind regime. Apart from the initially faster rise the subsequent evolution of the mass-loss rate is quite similar, with a pronounced maximum followed by a decline with decreasing luminosity and HBB efficiency as the envelope is removed. The production factors for this model are shown as error bars in Fig.~\ref{fig-yield}.

The almost halving of the duration of the TP-(S)AGB phase results in a lower yield of some 3DU (\chem{25}Mg, \chem{26}Mg) and HBB (\chem{4}He, \chem{27}Al) products. One would expect this model, with a TP-(S)AGB duration intermediate between the VW93 and B95 fixed composition opacities models, to also have yields lying between these results. This is mainly true but for a few exceptions mainly the yields of C, Na and F which we explain below.  

\begin{figure}
 \resizebox{\hsize}{!}{\includegraphics{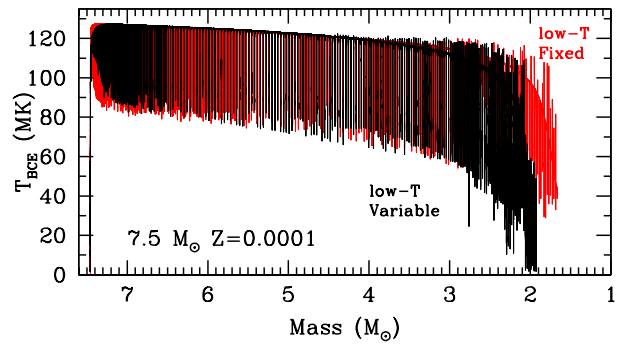}}
 \caption{Temperature at the base of the convective envelope as a function of total stellar mass for the 7.5\,M$_\odot$ Z=0.0001 models with VW93 mass-loss rate with fixed (red) and variable (black) molecular low temperature opacities.}
 \label{fig-tbce2}

 \end{figure}

As seen in Fig.~\ref{fig-tbce2} the temperature at the base of the convective envelope in the variable low-T molecular opacity model decreases near the end of the evolution when the envelope mass is higher compared to the fixed low-T molecular opacity model. 
This leads to two important consequences. First, the TP-SAGB phase terminates earlier in the variable opacity model because the instability which causes convergence issues is generally encountered only when HBB has ceased or is greatly reduced in efficiency. Secondly, the effect of this earlier cessation of HBB results in a higher yield of the 3DU products which are highly fragile to H burning namely \chem{12}C and \chem{19}F. The effect on the yield of \chem{23}Na however, is more complicated, with the variable low-T molecular opacity model case lying below both the fixed composition low-T opacity VW93 and B95 models. This is due to the \chem{23}Na having been depleted from HBB rapidly, then the majority of mass being lost at times of low abundance, with the larger \chem{23}Na near the end of the evolution not being able to overcome this previous release of Na poor material.

Because of the significant disparity between yield results obtained with/without variable composition molecular opacities the 6.5\,M$_\odot$ and 7.0\,M$_\odot$ Z=0.0001 VW93 mass-loss rate models were also recalculated using these updated opacities. A similarly large decrease in the duration of the TP-(S)AGB phase and number of thermal pulses was found (Table~\ref{table1}). The yield results for these models also echoed the trends of the 7.5\,M$_\odot$ Z=0.0001 model, with less 3DU enrichment and more moderate HBB nucleosynthesis. In these two lower mass models, the \chem{23}Na yields lie between the previous fixed composition VW93 and B95 models.  This highlights the significant interplay between the competing effects of mass-loss and third dredge-up near the end of the evolution that make yield calculation, especially of this isotope, quite sensitive.

We note that although low-T variable composition molecular opacities are an important and appropriate piece of input physics to be used in stellar models, the factor of two decrease in duration of the TP-(S)AGB with their inclusion in this work has far less of an effect than the factor of 10 difference seen from using different commonly used mass-loss rate prescriptions.
This impact on the yields is highlighted in the bottom panel of Fig.~\ref{fig-yield}.

As seen in the previous section, the nucleosynthesis for super-AGB models of different mass-loss rates is similar until near the end of the evolution. This weak dependence on the mass-loss rate is due to the very efficient hot bottom burning which is relatively constant for the majority of the TP-(S)AGB phase. For this reason, our initial, fixed low temperature opacity models have been left in this work as a complementary set to highlight/mimic the behaviour of a slower (non VW93) mass-loss rate. 

Mass-loss formalisms that are highly dependent on luminosity, e.g. B95, are far less affected by inclusion of variable composition opacities than prescriptions, such as VW93, which have a large dependence on stellar radius \citep{ven10b}. For this reason we decided not to recompute the B95 mass-loss rate models with low-T variable composition molecular opacities.
Comparing the yields of the variable composition opacities VW93 models with the B95 models, we find closer agreement than with the fixed opacity case, but significant differences still remain. Instead of the $\approx$ 2 orders of magnitude variations in yields of \chem{22}Ne, \chem{23}Na, \chem{24,25,26}Mg and \chem{27}Al, these have been slightly reduced to factors of $10-60$ variations in yields of the above listed species.   

The 7.0\,M$_\odot$ Z=0.001 model was recomputed using the updated opacities, with the evolution found to be reasonably comparable, with an approximately 20 per cent decrease in the number of thermal pulses and duration of the TP-(S)AGB phase. With such a modest change we have decided not to recompute the yield results for the Z=0.001 models.

 \subsubsection{Extrapolated thermal pulses}\label{subsec-extrap}

 Lastly we briefly discuss the effect of extrapolated thermal pulses at the end the evolution on the computation of yields. These extrapolations were performed to account for possible further evolution after convergence difficulties cease calculations. For the B95 mass-loss rates cases there are at most 2 further extrapolated pulses, whilst for VW93 mass-loss rate there are between 5-30 further thermal pulses (see Table~\ref{table3} in Appendix for extrapolated thermal pulse characteristics). As with our more metal rich models (Paper II) the isotopes most affected by extrapolated thermal pulses are the 3DU products \chem{12}C, \chem{19}F, \chem{21}Ne, \chem{26}Mg and $g$. Quite large variations are seen in \chem{19}F (0.4 dex) and \chem{21}Ne (0.8 dex) and more modest changes of up to $\sim$ 0.3 dex in \chem{12}C, 0.16 dex in \chem{26}Mg and 0.08 dex in $g$. At Z=0.0001 \chem{24}Mg and \chem{23}Na are also noticeably modified by these extrapolations, with up to 0.3 dex and 0.1 dex increase, respectively. However, since these extrapolated thermal pulses make up only a small fraction of the overall number of thermal pulses the nucleosynthetic effects from their inclusion are minimal for the majority of species (see columns 5-7 in online tables) and thus represent a small uncertainty compared to other effects, such as the mass-loss rate. 

 \subsection{Comparison to other studies: Z=0.0001}

Unlike solar metallicity models where super-AGB yield results in the literature show close agreement due to the unifying behaviour of HBB (e.g. see figure 15 in Paper II) this is not the case at lower metallicities. We compare the yield results of our representative 7.5\,M$_\odot$ Z=0.0001 (variable composition opacities) model with those of \citeauthor{sie10} (2010, hereafter S10). Even though both studies use the same mass-loss rate (VW93), the number of TPs and duration of the TP-(S)AGB phase vary considerably with 2776 TPs and $\tau_{\rm{(S)AGB}}$=3.56$\times$10$^{5}$ years in S10, compared to 248 TPs and $\tau_{\rm{(S)AGB}}$=6.23$\times$10$^{4}$ years in this work. 
These evolutionary differences are related to the feedback from 3DU events to the stellar structure as well as differences in core mass and its associated higher luminosity, with these factors helping drive more rapid mass loss. The difference in core mass is a result of different convective boundary treatment during core He burning. The S10 models were also calculated using the fixed composition opacities of \cite{fer05b}. As the core mass is an important factor in evolution and nucleosynthesis, we also compare to the 8.5\,M$_\odot$ Z=0.0001 model from S10 which at the beginning of the TP-SAGB phase has a more similar core mass to the 7.5\,M$_\odot$ Z=0.0001 model from this study. The structural properties for these models are presented in Table~\ref{compare2}. While the differences in duration of the thermally pulsing phase between our model and those from S10 are approximately a factor of $3-5$, the number of TPs varies by approximately a factor of 10. These differences are caused primarily by the lengthening of the interpulse period in models which undergo deep 3DU (e.g \citealt{sac77}), and hence our models have fewer TPs compared to those from S10 for the same thermally pulsing duration.
 
\begin{figure*}
\resizebox{\hsize}{!}{\includegraphics{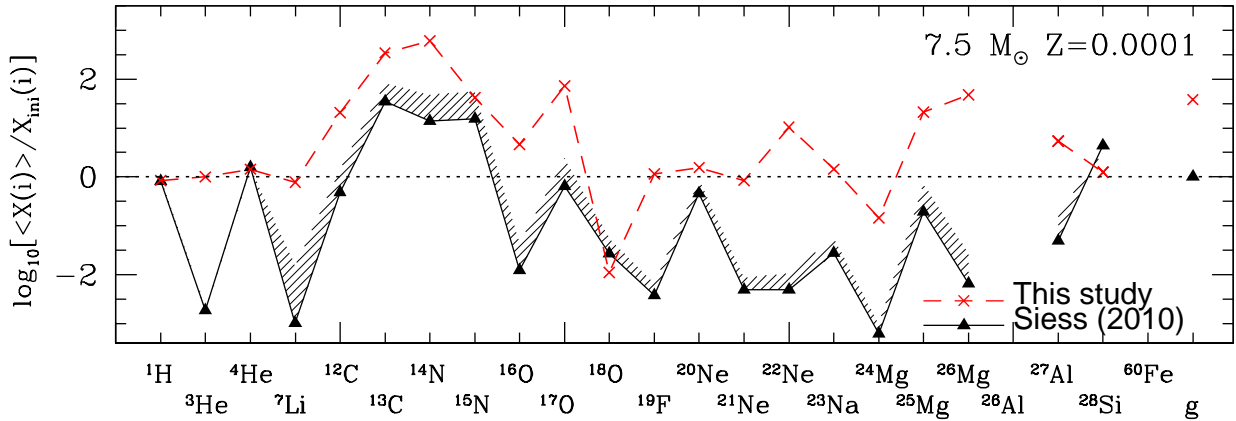}}
\caption{Production factor for the 7.5\,M$_\odot$ Z=0.0001 model from this study compared to that from \protect\cite{sie10}. The strong disparity between results is due mainly to different nuclear reaction rate networks, the large difference in temperature at the base of the convective envelope and the occurrence of 3DU in this work. The extent of the black shaded regions represents the yields from the 8.5\,M$_\odot$ Z=0.0001 model from S10. }
\label{fig-s10compare}
\end{figure*}

In Fig.~\ref{fig-s10compare} we provide the nucleosynthetic results as production factors for the three above-mentioned models. The shaded region represents the extent of the difference between the 7.5\,M$_\odot$ and 8.5\,M$_\odot$ Z=0.0001 models from S10, with the results from these models showing very close agreement.  
The yield divergences between our model and those from S10 are mainly due to the occurrence of 3DU in this current work, the large disparity in temperatures at the base of the convective envelope, and differences in the nuclear reaction rates. In particular, if the \chem{23}Na(p,$\gamma$)\chem{24}Mg reaction is much faster than the \chem{23}Na(p,$\alpha$)\chem{20}Ne then the Ne-Na cycle forms a chain and these isotopes flow through into the Mg-Al chain. Contrary to this work, the faster (p,$\gamma$) rate used in S10 (\texttt{NACRE}) leads to a substantial production of \chem{28}Si. 
In the S10 models, the 3DU products \chem{12}C, \chem{16}O, \chem{24,25,26}Mg are not replenished during the TP-(S)AGB phase. This makes the envelope abundances quite different and the contrast is exacerbated by the action of HBB.
The variation in yields between these studies is considerable, with many major elemental yields even varying from positive to negative. For example C, O, F, Ne, Na, Mg are all produced in our calculations whilst destroyed in the models of S10. 

We have not made comparison to the synthetic yield calculations in S10 which artificially model 3DU because they do not take into account the considerable feedback of these events on the structure which affects the resultant nucleosynthesis.

Comparison of results between different studies for Z=0.001 models can be found in the next section, as they are connected to the study of globular cluster abundance anomalies.

\begin{table}
\begin{center}\setlength{\tabcolsep}{1.5pt} 
\caption{Comparison of selected model characteristics for 7.5\,M$_\odot$ Z=0.0001 from this study and Siess (2010). Variables as described in Table~\ref{table1}.}
\label{compare2}
\begin{tabular}{lcccccrc} \hline \hline
&M$_{\rm{ini}}$&$T_{\rm{BCE}}^{\rm{Max}}$&$M_{\rm{Dredge}}^{\rm{Tot}}$ & $M_{\rm{2DU}}$ & $M_{\rm{C}}^{\rm{F}}$ &  $N_{\rm{TP}}$  & $\tau_{\rm{(S)AGB}}$ \\
&(M$_\odot$)&(MK)&(M$_\odot$)&(M$_\odot$)&(M$_\odot$)&&(yrs)\\\hline
This Study  &7.5&127&3.02(-2)&1.21 & 1.22 &248 & 6.23(4) \\
Siess (2010)&7.5&155&0       &1.07 & 1.28 &2776 & 3.56(5)\\
Siess (2010)&8.5&151&0       &1.20 & 1.30 &2409 & 1.64(5)\\
\hline
 \end{tabular} \end{center} \end{table}

\section{Globular cluster extreme population : NGC 2808}\label{sec-ngc2808}

\begin{figure*}\begin{center}
 \includegraphics[width=7cm,angle=0]{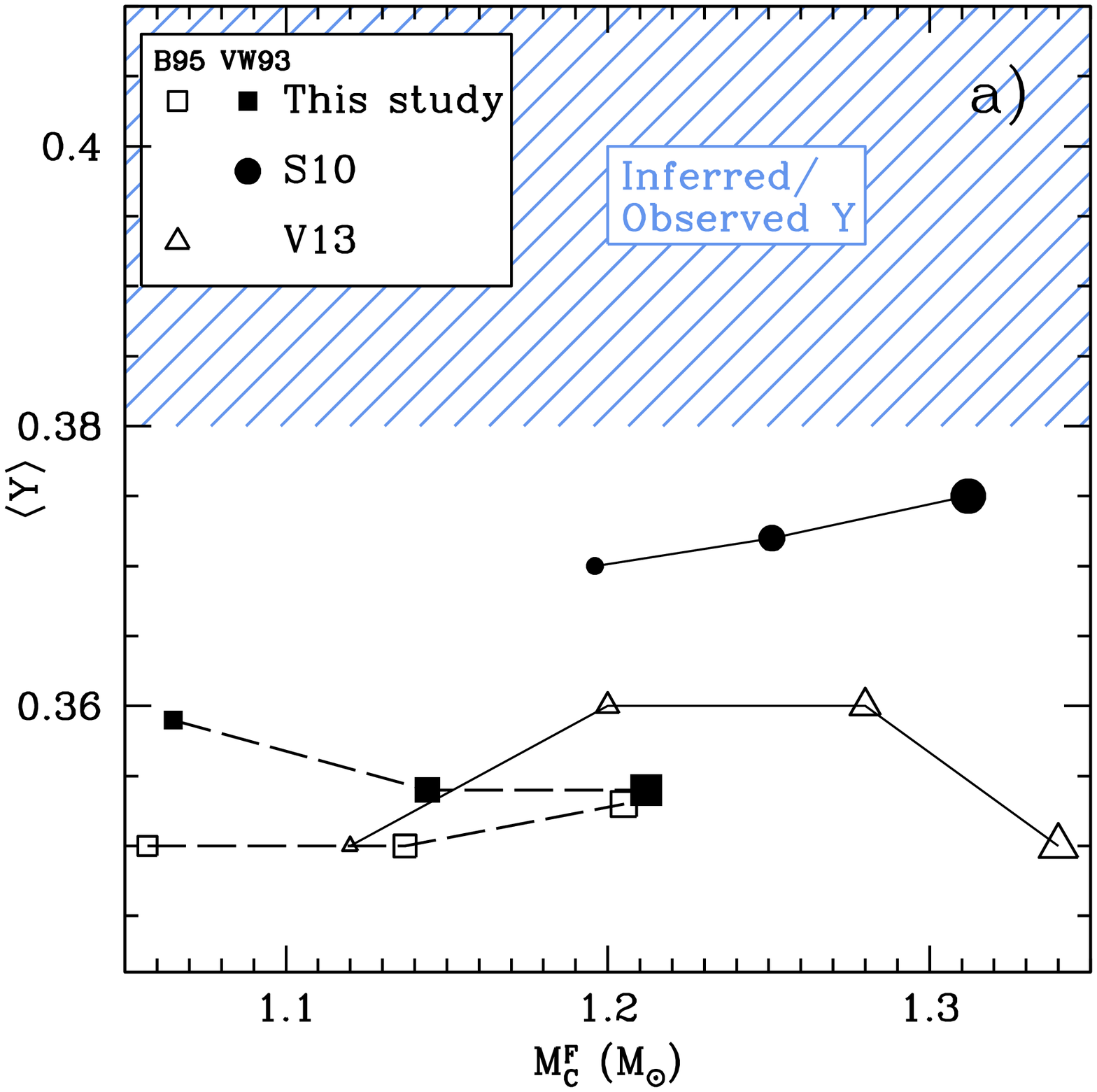}
 \includegraphics[width=7cm,angle=0]{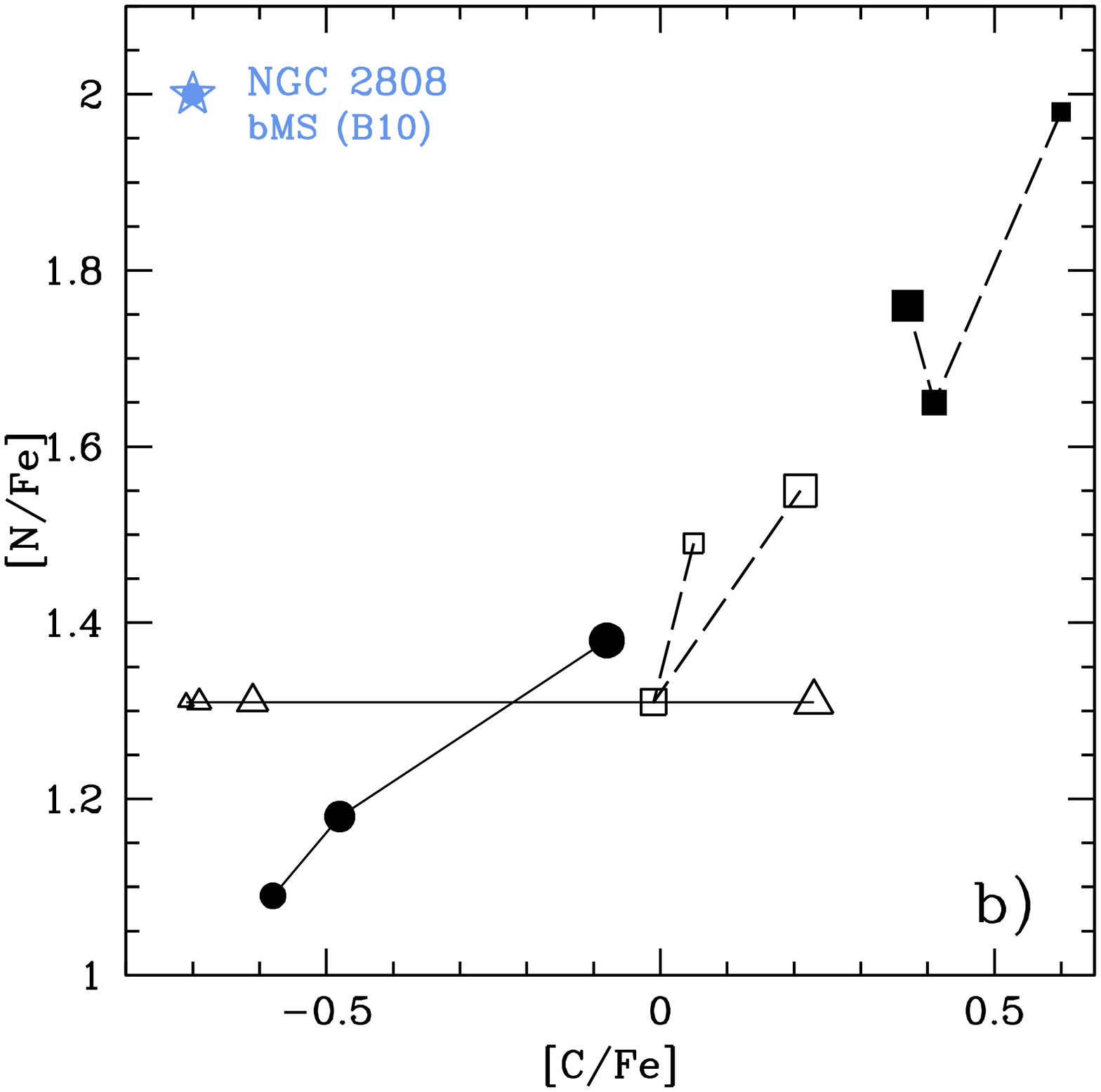}
 \includegraphics[width=7cm,angle=0]{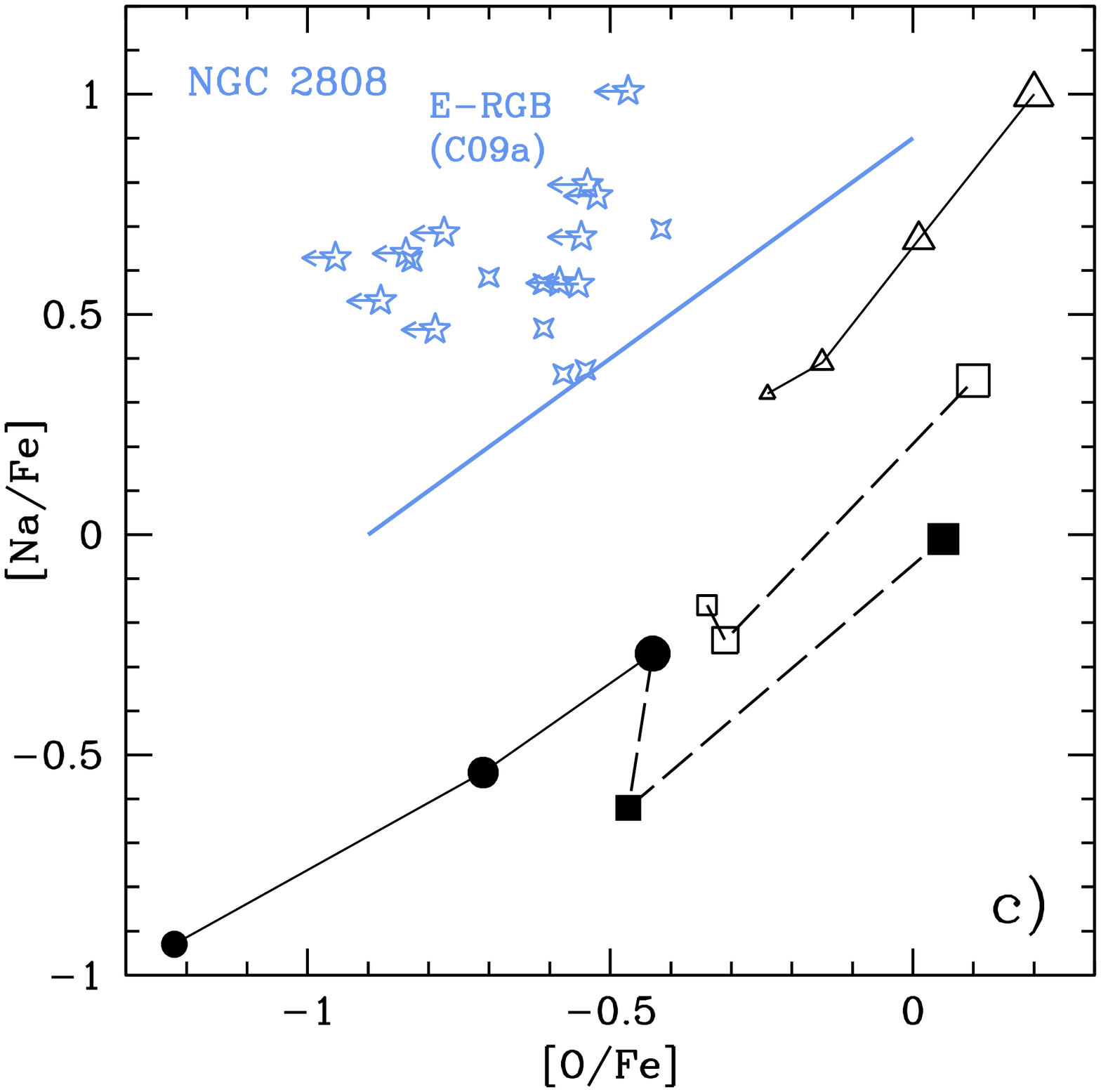}
 \includegraphics[width=7cm,angle=0]{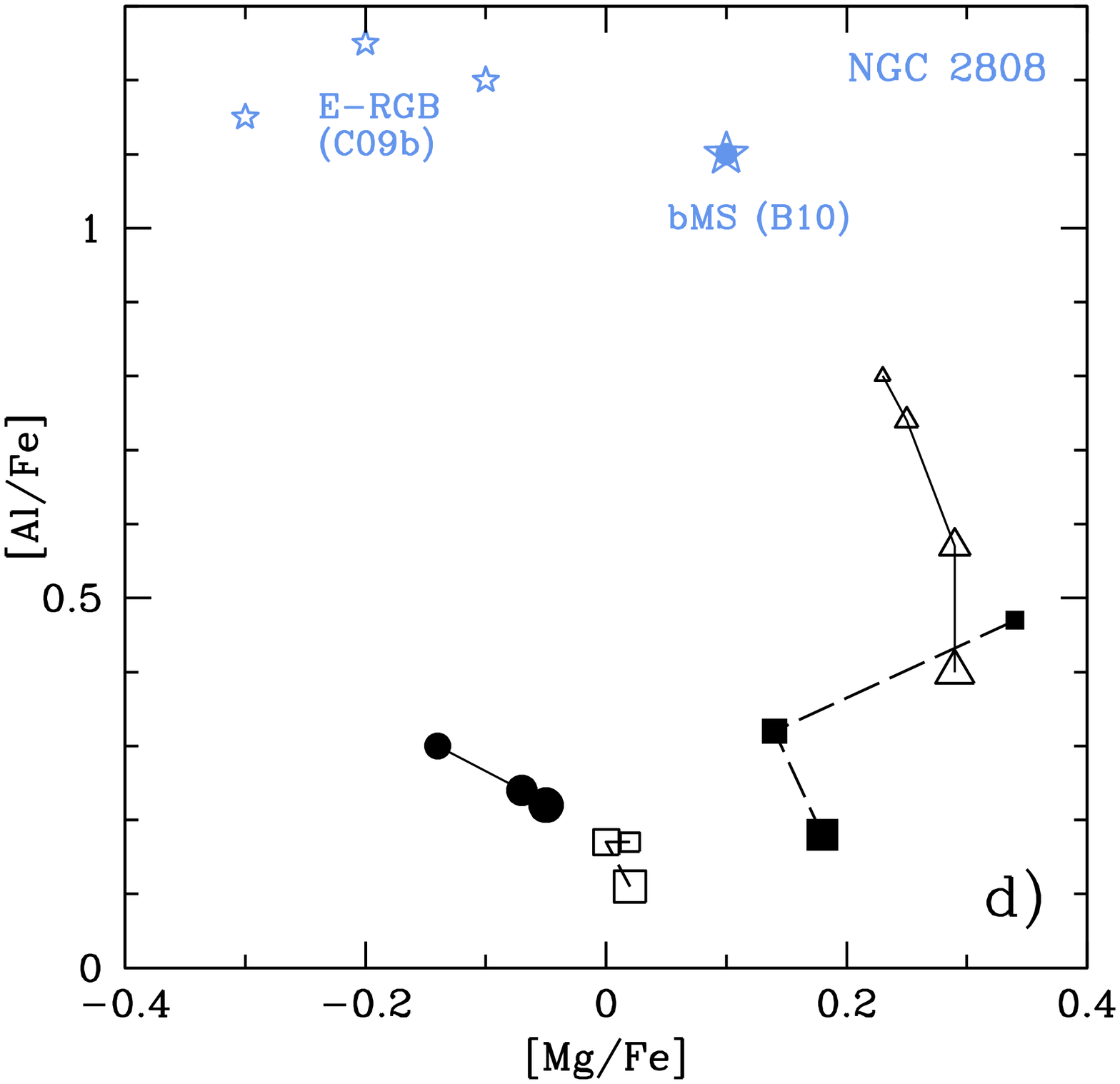}
 \caption{Comparison of model yield results to observations. The size of the symbols is proportional to mass, with larger symbols representing more massive models. The open symbols represent models using B95 mass-loss rate, whilst closed symbols are models computed using the VW93 mass-loss rate. The dashed line joins results from the current work, while solid lines connect results from the literature, also identified by the symbols. a) Average helium mass fraction in the wind $\langle$Y$\rangle$ against final core mass. The initial value for this work and S10 was 0.2492 but 0.24 in V13. The shaded region in this figure represents the inferred/observed helium abundance in the extreme population in NGC 2808. b) [N/Fe] versus [C/Fe], c) [Na/Fe] versus [O/Fe] where the solid blue line represents the E division from \protect\cite{car09a} d) [Al/Fe] versus [Mg/Fe]. Observations from NGC 2808 in panel c) are shown with open star symbols, with the symbols with arrows representing upper O limits from \protect\cite{car09a}, the large filled star symbol in panels b) and d) represents the blue MS star from \protect\cite{bra10a}.}
\label{fig-giant}  \end{center} \end{figure*}

\begin{figure*}\begin{center}
 \includegraphics[width=7cm,angle=0]{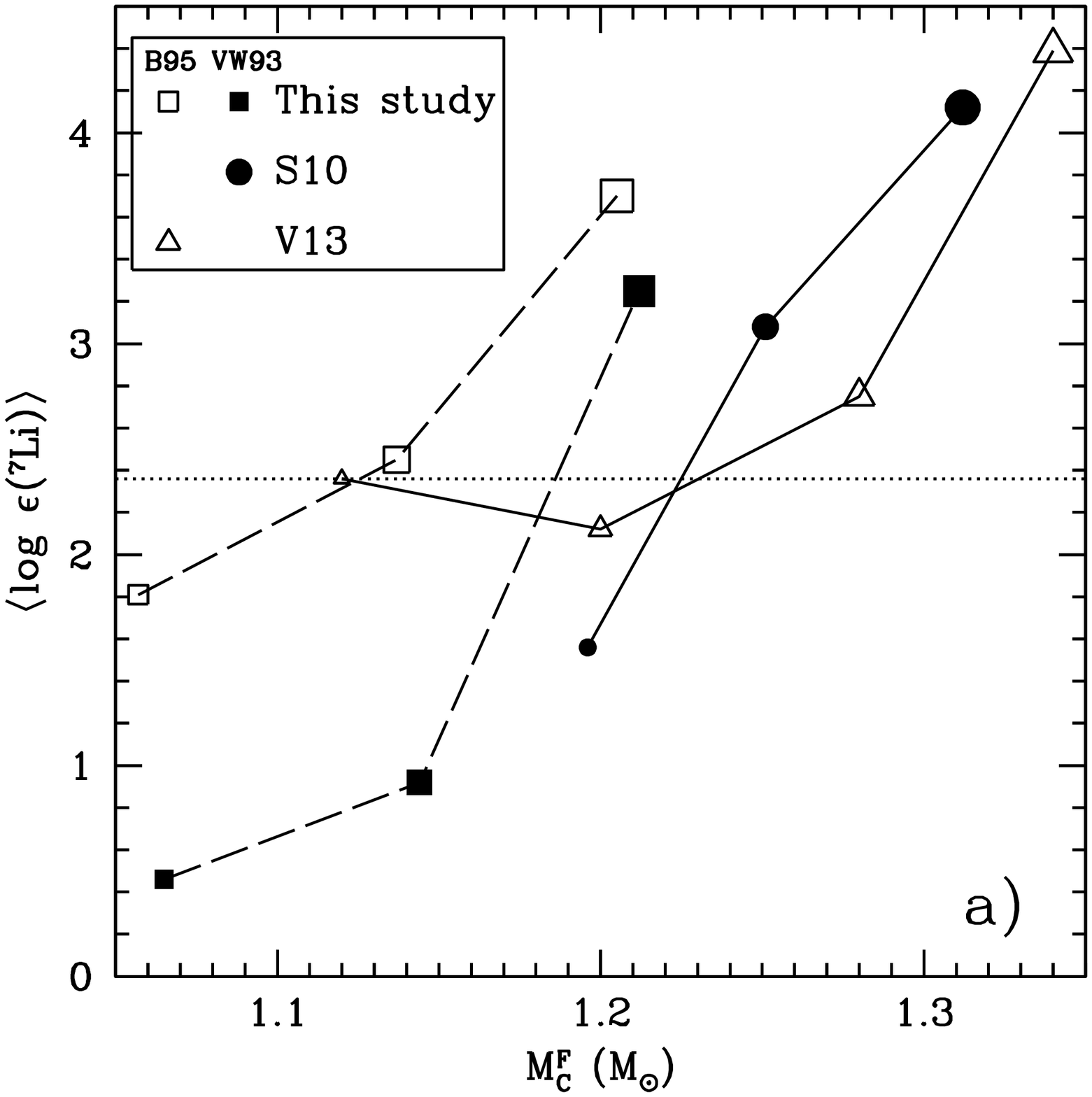}
 \includegraphics[width=7cm,angle=0]{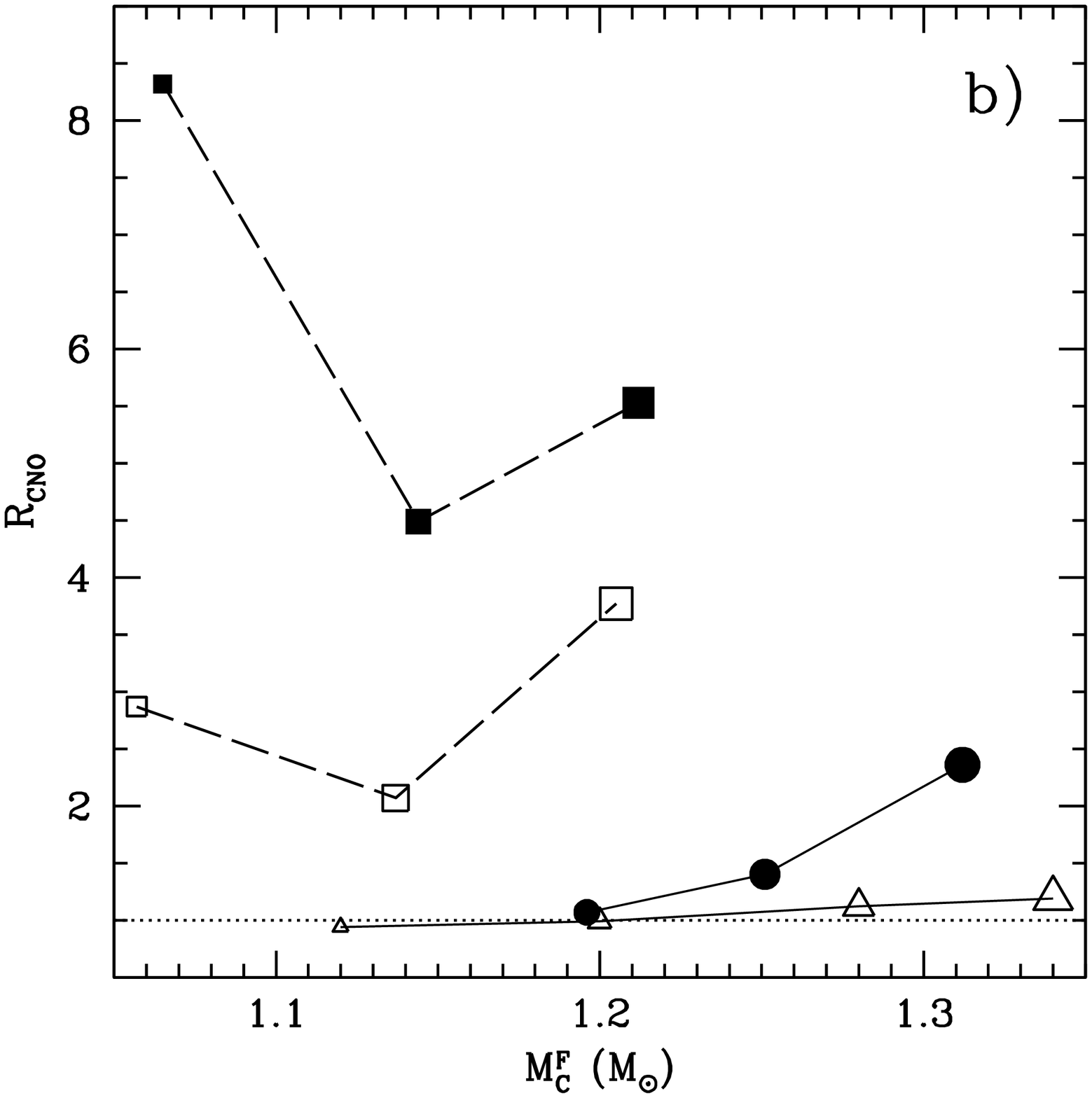}
 \caption{Theoretical predictions of a) Lithium yield as $\langle$log$_{10}$$\epsilon$(\chem{7}Li)$\rangle$ and b) $R_{\rm{CNO}}$ both plotted against final core mass. Symbols and lines as in Fig.~\ref{fig-giant}. The dotted horizontal lines in each panel represents the initial value of the respective quantity.}
\label{fig-giant2}  \end{center} \end{figure*}

As mentioned in the introduction, one of the main hypotheses to explain the extreme stellar population within globular clusters involves the stars being formed directly from pristine super-AGB ejecta \citep{der08,der12}. If this scenario holds, then the chemically anomalous patterns observed in the extreme population should therefore directly echo the super-AGB nucleosynthetic yields.
Using our computed yields of metallicity Z=0.001, as well as previous results from the literature, we aim to test the validity of this scenario by comparison with observational data from NGC 2808, one of the most thoroughly studied mono-metallic globular clusters which hosts an extreme population.

NGC 2808 is a massive cluster with metallicity [Fe/H]$\sim-$1.1 \cite[][2010 edition]{har96}.  A large helium enhancement between the primordial and intermediate or extreme populations is inferred from features such as the horizontal branch (HB) morphology \citep{dan04,lee05} and  the triple main sequence splitting \citep{pio07}. To match these features, the cluster is thought to harbour at least three\footnote{\cite{gra11} have recently shown that NGC 2808 is made up of \textit{more} than three separate populations, with the intermediate population further split into two.} distinct populations, with helium contents (Y) of $\sim$ 0.245, $\sim$ 0.30-0.33 and $\sim$ 0.38-41 \citep{dan05,lee05,dan07,pio07,bra10b}. The difference in helium abundance between populations $\Delta$Y  has been observationally estimated as 0.09 \citep{mar13} and 0.17 \citep{pas11} for the I-P and E-P populations respectively. 
 
Apart from the helium enhancement, this cluster also shows very extended C-N, O-Na, Mg-Al and Mg-Si anti-correlations (\citealt{car09a}, \citealt{car09b}, \citealt{bra10a}).
 
In Fig.~\ref{fig-giant} we draw together our nucleosynthetic yield results and compare them to the main observable (and inferred) abundances within the extreme population of NGC 2808; we focus on \chem{4}He and the anti-correlations C-N, O-Na and Mg-Al. In Fig.~\ref{fig-giant2} we also highlight the yield of \chem{7}Li and $R_{\rm{CNO}}$, noting that observations of these quantities are currently not available for NGC 2808. 

From a theoretical perspective we also add to these figures yield results from S10 and \citeauthor{ven13} (2013, hereafter V13). The main input physics used in these studies can be briefly summarised as follows.
The S10 models are characterised by convective mixing using MLT with $\alpha_{\rm{MLT}}$=1.75, moderate mass-loss (VW93) and scaled solar composition. The models of V13 have efficient convective mixing using the FST prescription, rapid mass-loss (B95 $\eta$=0.02) and $\alpha$ enhanced composition with +0.4 in [O/Fe], [Mg/Fe] and [Si/Fe].

In this study, and that of V13, some form of overshoot was used during the core He burning phase which results in a similar range of initial masses for super-AGB stars ($\sim$ $6.5-8.0$\,M$_\odot$)\footnote{The \textsc{monstar} evolution code uses the search for convective neutrality approach of \cite{lat86} to determine convective boundaries. In the \textsc{aton} program convective overshooting was employed during both core H and He burning using a exponential decay of convective velocities starting from convective boundaries, with the e-folding distance given by $\zeta$$H_P$ with $\zeta$ =0.02 \citep{ven13}}. The S10 models, with their use of a strict Schwarzschild criterion to delineate convective boundaries, have a smaller core mass at the beginning of the TP-(S)AGB phase for the same initial mass, and as a result the super-AGB star mass range is translated upwards by $\sim$ 1\,M$_\odot$. As the core mass is the key driver of further evolution we make our comparisons as a function of core mass $M_{\rm{C}}$ instead of initial mass $M_{\rm{ini}}$.

The efficiency of 3DU varies widely in computational studies of low metallicity super-AGB stars, with either no 3DU ($\lambda$ = 0) in V13 and S10, moderate 3DU ($\lambda$$\sim$0.7-0.8 - this study) or very efficient 3DU ($\lambda$ $\la$ 1) in \cite{her12}\footnote{The \cite{her12} models are of slightly lower metallicity than discussed here (Z = 0.0006), with the study focusing on the early TP-(S)AGB, so unfortunately we cannot compare stellar yield predictions.}. In low-mass AGB star computations the efficiency of 3DU increases with decreasing metallicity \cite[e.g.][]{woo81c,str06}. For super-AGB stars models however, we find that 3DU efficiency is practically independent of initial metallicity (at least for 0.02 $\la$ Z $\la$ 0.0001) and correlated primarily with core mass (Paper V - Doherty et al., in preparation). 

The nucleosynthesis of super-AGB stars is affected by the duration of the TP-(S)AGB phase, and number of thermal pulses, with these quantities varying considerably between studies, ranging from $\sim$ 5800-16000 years (28-32 TPs) in V13, $\sim$ 88000-117000 years (394-1237 TPs) in S10, and $\sim$ 14300-169000 years (28-171 TPs) in the current work. 

The observational data for NGC 2808 were taken from a selection of studies; Mg and Al abundances from red giant branch (RGB) stars using high resolution UVES spectra \cite[][hereafter C09b]{car09b}, O and Na results also for RGB stars taken with GIRAFFE \cite[][hereafter C09a]{car09a} whilst \citeauthor{bra10a} (2010, hereafter B10) using X-SHOOTER observed a selection of elements C, N, Na, Mg and Al in two main sequence (MS) stars, one from the blue (suspected extreme population) MS and the other from the red (suspected primordial population) MS. Here we make comparison to a small subset of these observational results, focusing on those stars which lie in the extreme population.

\subsection{Light elements : He and Li}

Fig.~\ref{fig-giant}a shows the average mass fraction of helium in the wind $\langle$Y$\rangle$ as a function of final core mass for all model results. The shaded region $\langle$Y$\rangle$ $\sim$ 0.38-0.41 corresponds to the inferred/suspected \chem{4}He content of the extreme population stars within NGC 2808.
As we have explained in Section~\ref{sec-nucleo} the contribution from 3DU and HBB to \chem{4}He production is not a major factor compared to 2DU in this mass and metallicity range. The S10 models show the greatest enrichment, primarily due to the use of strict Schwarzschild criterion for convective boundaries which leads to a smaller core during central helium burning and a thicker He-rich layer, which the base of the convective envelope can penetrate during 2DU. Although all models show considerable enrichment of helium from 2DU, we agree with previous studies that the extreme population helium contents cannot be reached with current super-AGB star theoretical predictions.

The behaviour of the lithium yield as a function of final core mass can be seen in Fig.~\ref{fig-giant2}a. Whilst there are large variations between model results, the trend is for increasing yield with increasing initial mass and/or mass-loss rate, due to the more rapid expulsion of the envelope during the lithium-rich phase early on the AGB.
As yet no \chem{7}Li abundances have been measured in NGC 2808 stars, however, if super-AGB stars (in particular the most massive super-AGB stars) are indeed the source of the extreme population material then a testable consequence should be that \chem{7}Li is significantly enhanced in the extreme population \citep{dan12}.

\subsection{CN and C+N+O}

Fig.~\ref{fig-giant2}b shows the $R_{\rm{CNO}}$ versus final core mass for the selection of theoretical yield predictions. 

Prior to the TP-(S)AGB phase there are two processes whereby the total surface $R_{\rm{CNO}}$ can be increased. The first, corrosive 2DU which was explored in detail in Section~\ref{sec-2du} results in the addition of primary \chem{12}C (and \chem{16}O) to the surface. The second process, a dredge-out event \citep{pap4,sie06}, occurs in the most massive super-AGB stars near the end of the carbon burning phase. These events are characterised by the formation of a convective helium burning region which then merges with the incoming convective envelope. A substantial amount of helium processed material (primarily carbon) is dredged to the surface. 
We find that, irrespective of the occurrence or not of 3DU, due to C2DU or dredge-out events our models with core masses $M_{\rm{C}}$ $>$ 1.18\,M$_\odot$ show a substantial (greater than a factor or two) increase in $R_{\rm{CNO}}$. This amount of C+N+O surface enrichment is similar in magnitude to the results obtained in rotating intermediate-mass star models by \cite{dec09}.
\cite{pum08} suggested that \textit{all} super-AGB stars that experience a dredge-out event must then go through an electron capture supernova event to rid the cluster of this highly CNO enriched material. The slow mass-loss rate required to access the supernova channel however, is at odds with the premise that rapid mass-loss rates are required for the slightly less massive (S)AGB models to provide the extreme population material. Since the carbon content, as well as the total metallicity, drive structural changes within AGB stars, the level of enrichment from C2DU/dredge-out is important to the subsequent evolution. These phenomena and the factors that affect them are currently under investigation.

We note here that the most massive V13 models show either a slight decrease or only minimal increase in $R_{\rm{CNO}}$ which suggests these models do not undergo a C2DU or a dredge-out event. 

Unfortunately the C+N+O abundance of the stellar populations has not been measured in NGC 2808, but if the expected constancy between populations is observationally confirmed, and if the polluters were indeed super-AGB stars, this would require them to have no or inefficient 3DU, and/or a very rapid mass-loss as well as undergo no corrosive 2DU or dredge-out events.

Fig.~\ref{fig-giant}b shows the theoretical yield predictions of [C/Fe] and [N/Fe] as well as the observational data for the extreme population blue MS star from the B10 study. All super-AGB star model results show a large N enhancement with [N/Fe] values of between 1-2 from CNO cycling during HBB, with the CN reaching its equilibrium value. Since the S10/V13 models lack 3DU this translates to a lower N and C abundance.
There is a substantial spread in results using different mass-loss rates as well as when compared to the previous studies. All our standard models show a C-N correlation, with the large increase of C from the contributions of corrosive 2DU and/or 3DU. The blue MS B10 star shows a particularly high N and low C value, which is not matched by any of the model predictions.  

\subsection{O-Na}

The O-Na anti-correlation is a ubiquitous feature of globular clusters to various extents \citep{car09a}.  However, it is a well-established result that O and Na abundances become correlated as a result of very hot hydrogen burning \citep{den03} and this is confirmed in all super-AGB models. 
Fig.~\ref{fig-giant}c shows [O/Fe] vs [Na/Fe], with the observational results of C09a and the 
(arbitrary) dividing line between I and E populations which corresponds to [O/Na] $>$ $-0.9$ (C09a)

Lower mass models have the lowest [O/Fe] values, due to more ON cycling during the longer TP-(S)AGB phase. Our most massive model of each mass-loss rate has positive oxygen yields due to the introduction of primary oxygen from CO2DU. 
The vast depletion of oxygen required to match the extreme population is achieved in the lowest mass model of S10, although with large sodium destruction. 
Taking into account an enhanced initial O abundance ([O/Fe]=0.4), the lowest mass V13 model has depleted at most $\sim$ 0.7 in [O/Fe]. Clearly this amount of depletion is not enough to match the extreme population. To further decrease O it has been suggested \cite[e.g.][]{dan07,dan11,der12} that some unknown extra-mixing must (1) only take place within the extreme population of stars (2) only operate during the RGB phase and (3) work to deplete O but without affecting the Na content. We note that the magnitude of O depletion required to match the extreme population in NGC 2808 would have to be quite significant, of the order $\sim$ $0.7-1.0$ in [O/Fe]. Unfortunately, there are no O observational determinations from an extreme population MS star in NGC 2808. If a large depletion of O is found to be already present within stars at this early evolutionary stage it would clearly rule out the extra-mixing scenario.

In super-AGB stars \chem{23}Na is primarily a 2DU product, which is then subsequently depleted via HBB. 
To achieve the largest \chem{23}Na production, the mass-loss rate has be very rapid and the \chem{22}Ne(p,$\gamma$)\chem{23}Na reaction must out-compete the destruction channels \chem{23}Na(p,$\alpha$)\chem{20}Ne and \chem{23}Na(p,$\gamma$)\chem{24}Mg (Fig.~\ref{fig-nena}). A larger initial \chem{20}Ne abundance \citep{der10} or increased \chem{20}Ne(p,$\gamma$)\chem{21}Na reaction rates \citep{der10} also have some, albeit not substantial, effect on larger \chem{23}Na yields.

\subsection{Mg-Al, Si}

The most obvious feature in Fig.~\ref{fig-giant}d, which shows [Mg/Fe] vs. [Al/Fe], is the great disparity between the model results and the observations. No theoretical super-AGB star predictions are able to reproduce the simultaneous large Al production and Mg destruction for the 3 RGB stars from C09b. We note here that even if any form of mixing had taken place prior to or during the RGB phase, the Mg and Al would not have been altered, because the quite large temperatures necessary to modify these elements are not achieved within these relatively low mass stars \citep{den98}.

When considering only HBB nucleosynthesis, higher temperatures at the base of the convective envelope as well as a longer TP-(S)AGB lifetime (i.e. lower mass-loss rates) both result in more Mg depletion and Al production. The largest Mg depletion and Al production require that \chem{25}Mg, which is mostly produced after proton capture on \chem{24}Mg, is efficiently depleted by \chem{25}Mg(p,$\gamma$)\chem{26}Al followed by \chem{26}Al(p,$\gamma$)\chem{27}Si($\beta^{+}$)\chem{27}Al to bypass the \chem{26}Mg production (Fig~\ref{fig-nena}). However, when $T_{\rm{BCE}}$ exceeds $\sim$ 120-130 MK leakage from the Mg-Al chain via \chem{27}Al(p,$\gamma$)\chem{28}Si starts depleting Al (Fig.~\ref{fig-nena}).

The models of S10 and V13 show a similar behaviour of increasing Al production and Mg destruction with decreasing initial mass. The depletion of magnesium in the V13 models is slight, at about 0.11-0.17 from the original value of [Mg/Fe]= 0.4. The S10 models show quite a similar depletion of [Mg/Fe]$\sim$0.1-0.2 mainly attributed to the longer TP-(S)AGB duration in these models. The large Al production in V13 is due to their use of the upper \texttt{NACRE} limit for \chem{25}Mg(p,$\gamma$). In \cite{ven11b} a test using a massive AGB model (6.0\,M$_\odot$) was performed whereby the \chem{25}Mg(p,$\gamma$)\chem{26}Al reaction rate was doubled\footnote{The recent measurements of the \chem{25}Mg(p,$\gamma$)\chem{26}Al (ground state and meta stable) reactions  by \cite{str13} do confirm that the current best estimate of this rate is close to that proposed by \cite{ven11b}} to maximize Mg depletion to try to match the three Mg-poor stars in the C09a study. In this model they found modest [Mg/Fe] depletion to 0.13 and [Al/Fe] value of 1.18 which does a good job of reproducing the B10 star. However, there is still a need for an additional depletion of [Mg/Fe] by 0.4 to reach the C09b observations. Our VW93 models show an increase in [Mg/Fe] due to efficient 3DU of \chem{25}Mg and \chem{26}Mg. The B95 models with their shorter duration show very little Al production and due to smaller 3DU contribution maintain [Mg/Fe] values close to the origin.

In \cite{car09b} a slight Al-Si correlation, with slope of 0.08 was uncovered within NGC 2808. The amount of silicon production in S10, V13 and this current study are modest, slightly less than required to match this correlation.

\section{Summary and Conclusions}\label{sec-conclude}

We have computed a grid of metal-poor and very metal-poor super-AGB stars to explore element production. These stars create large amounts of \chem{4}He, \chem{13}C, \chem{14}N and \chem{17}O, as well as the heavy magnesium isotopes \chem{25}Mg and \chem{26}Mg. In addition, and contrary to higher metallicity models, we also find positive yields of \chem{12}C, \chem{15}N, \chem{16}O, and the heavier proton chain species \chem{27}Al and \chem{28}Si.  

The occurrence of third dredge-up in our models is a key difference compared to previous low metallicity super-AGB yield studies. Whilst there is evidence for third dredge-up in intermediate-mass/massive AGB stars of higher metallicities (e.g. Rb observations - \citealt{gar06,gar09}, very luminous C-stars - \citealt{van99}, large N overabundance - \citealt{mcs07}), at the metallicities considered here, the efficiency, or even the occurrence of 3DU is unknown.

Contrary to the higher metallicity models where the surface composition is driven almost purely by HBB, the pollution from dredge-up events plays an important role in our metal-poor and very metal-poor models. The dominant dredge-up process to affect the surface enrichment changes from 3DU at $\sim$ 6.5\,M$_\odot$, to about equal contribution from C2DU and 3DU at $\sim$ 7.0\,M$_\odot$ whilst the abundances in the more massive models are mainly dictated by the large enrichment prior to the TP-(S)AGB phase from CO2DU/dredge-out. The envelope enrichment in carbon from either corrosive 2DU or a dredge-out event prior to the TP-(S)AGB phase can have a large impact on the further evolution, primarily though enhanced mass loss driven via enhanced opacity.

We have explored a selection of uncertainties within our super-AGB star models. The mass-loss rate has the greatest impact with up to a factor of 10 difference in TP-(S)AGB lifetime of very metal-poor stars when two commonly used mass-loss rate prescriptions are used. The yields of isotopes most affected by changes to the mass-loss prescriptions are \chem{22}Ne, \chem{23}Na and \chem{24,25,26}Mg and \chem{27}Al and $g$. 
The appropriate treatment of low temperature molecular opacities is crucial to super-AGB star evolution for metallicities below Z$\approx$0.001, where they truncate the evolution by approximately 50 per cent. However, the inclusion of these updated opacities has a less dramatic effect than changes to the mass-loss rate prescription. A modest increase in $\alpha_{\rm{mlt}}$ teamed with a rapid mass-loss rate was found to have only a small impact on super-AGB star nucleosynthesis.
Extrapolated thermal pulses at the end of the evolution, to account for possible nucleosynthesis after convergence issues cease calculations, have negligible effect on the yields of most isotopes.

With the yields showing \textit{such} large variations both within, and between, different research groups, which observable nucleosynthetic signature of low metallicity super-AGB stars can be considered robust?
As the CNO elemental yields are dependent on 3DU efficiency and mass-loss prescription they show no consistency between code results. 
Lithium is notoriously temperamental, and is highly dependent on the mass-loss rate and the treatment of convective mixing. 
The Ne-Na cycle and Mg-Al chain isotope yields vary widely due to uncertainties in reaction rates, mass-loss prescriptions and occurrence of 3DU. Also with large variation in 3DU efficiency between calculations (0 $\la$ $\lambda$ $\la$ 1), all elements heavier than iron will vary considerable between calculations.
This leaves the large helium enhancement from 2DU, as well as certain isotopic ratios, such as \chem{12}C/\chem{13}C and \chem{14}N/\chem{15}N which reach their equilibrium values as the only consistent result between different studies. However, these model independent characteristics are not unique to super-AGB stars but are also found in intermediate-mass and massive AGB stars.

Whilst super-AGB are thought of as one of the most likely candidate polluters of the extreme population within NGC 2808, Fig.~\ref{fig-giant} shows that the observational data for this cluster and theoretical super-AGB yield predictions are not compatible. For the super-AGB pollution scenario to be salvaged extra-mixing within the extreme population needs to be invoked, and some observational results would have to be questioned. However, given the large quantitative uncertainties in both the theoretical and observational results we cannot completely rule out super-AGB stars.  

\section{Acknowledgments}

This research was supported under Australian Research Council's Discovery Projects funding scheme (project numbers DP0877317 and DP120101815), and the Go8-DAAD-Australia/Germany Joint research cooperation scheme. PGP was supported by the project AYA201233938. S.W.C acknowledges support from the Australian Research Council's Discovery Projects funding scheme (project number DP1095368). CLD would like to thank Valentina D'Orazi and Tom Constantino for helpful discussions. We would also like to thank the referee Onno Pols for his careful reading of the manuscript and helpful suggestions.

\bibliographystyle{mn2e}

\bibliography{P3doh}

\begin{thebibliography}{}

\bibitem[\protect\citeauthoryear{{Arnould}, {Goriely} \& {Jorissen}}{{Arnould}
  et~al.}{1999}]{arn99}
{Arnould} M.,  {Goriely} S.,    {Jorissen} A.,  1999, \aap, 347, 572

\bibitem[\protect\citeauthoryear{{Bekki}, {Campbell}, {Lattanzio} \&
  {Norris}}{{Bekki} et~al.}{2007}]{bek07}
{Bekki} K.,  {Campbell} S.~W.,  {Lattanzio} J.~C.,    {Norris} J.~E.,  2007,
  \mnras, 377, 335

\bibitem[\protect\citeauthoryear{{Bloecker}}{{Bloecker}}{1995}]{blo95}
{Bloecker} T.,  1995, \aap, 297, 727

\bibitem[\protect\citeauthoryear{{Bowen}}{{Bowen}}{1988}]{bow88}
{Bowen} G.~H.,  1988, \apj, 329, 299

\bibitem[\protect\citeauthoryear{{Bragaglia}, {Carretta}, {Gratton}, {D'Orazi},
  {Cassisi} \& {Lucatello}}{{Bragaglia} et~al.}{2010}]{bra10b}
{Bragaglia} A.,  {Carretta} E.,  {Gratton} R.,  {D'Orazi} V.,  {Cassisi} S.,
  {Lucatello} S.,  2010, \aap, 519, A60

\bibitem[\protect\citeauthoryear{{Bragaglia}, {Carretta}, {Gratton},
  {Lucatello}, {Milone}, {Piotto}, {D'Orazi}, {Cassisi}, {Sneden} \&
  {Bedin}}{{Bragaglia} et~al.}{2010}]{bra10a}
{Bragaglia} A.,  {Carretta} E.,  {Gratton} R.~G.,  {Lucatello} S.,  {Milone}
  A.,  {Piotto} G.,  {D'Orazi} V.,  {Cassisi} S.,  {Sneden} C.,    {Bedin}
  L.~R.,  2010, \apjl, 720, L41

\bibitem[\protect\citeauthoryear{{Campbell} \& {Lattanzio}}{{Campbell} \&
  {Lattanzio}}{2008}]{cam08}
{Campbell} S.~W.,  {Lattanzio} J.~C.,  2008, \aap, 490, 769

\bibitem[\protect\citeauthoryear{{Cannon}}{{Cannon}}{1993}]{can93}
{Cannon} R.~C.,  1993, \mnras, 263, 817

\bibitem[\protect\citeauthoryear{{Canuto} \& {Mazzitelli}}{{Canuto} \&
  {Mazzitelli}}{1991}]{can91}
{Canuto} V.~M.,  {Mazzitelli} I.,  1991, \apj, 370, 295

\bibitem[\protect\citeauthoryear{{Carretta}, {Bragaglia}, {Gratton} \&
  {Lucatello}}{{Carretta} et~al.}{2009}]{car09b}
{Carretta} E.,  {Bragaglia} A.,  {Gratton} R.,    {Lucatello} S.,  2009, \aap,
  505, 139 (C09b)

\bibitem[\protect\citeauthoryear{{Carretta}, {Bragaglia}, {Gratton},
  {Lucatello}, {Catanzaro}, {Leone}, {Bellazzini} \& {et.al.}}{{Carretta}
  et~al.}{2009}]{car09a}
{Carretta} E.,  {Bragaglia} A.,  {Gratton} R.~G.,  {Lucatello} S.,  {Catanzaro}
  G.,  {Leone} F.,  {Bellazzini} M.,    {et.al.} 2009, \aap, 505, 117 (C09a)

\bibitem[\protect\citeauthoryear{{Caughlan} \& {Fowler}}{{Caughlan} \&
  {Fowler}}{1988}]{cf88}
{Caughlan} G.~R.,  {Fowler} W.~A.,  1988, Atomic Data and Nuclear Data Tables,
  40, 283

\bibitem[\protect\citeauthoryear{{Cottrell} \& {Da Costa}}{{Cottrell} \& {Da
  Costa}}{1981}]{cot81}
{Cottrell} P.~L.,  {Da Costa} G.~S.,  1981, \apjl, 245, L79

\bibitem[\protect\citeauthoryear{{Cristallo}, {Straniero}, {Gallino},
  {Piersanti}, {Dom{\'{\i}}nguez} \& {Lederer}}{{Cristallo}
  et~al.}{2009}]{cri09}
{Cristallo} S.,  {Straniero} O.,  {Gallino} R.,  {Piersanti} L.,
  {Dom{\'{\i}}nguez} I.,    {Lederer} M.~T.,  2009, \apj, 696, 797

\bibitem[\protect\citeauthoryear{{Cyburt}, {Amthor}, {Ferguson}, {Meisel},
  {Smith}, {Warren}, {Heger}, {Hoffman}, {Rauscher}, {Sakharuk}, {Schatz},
  {Thielemann} \& {Wiescher}}{{Cyburt} et~al.}{2010}]{jin10}
{Cyburt} R.~H.,  {Amthor} A.~M.,  {Ferguson} R.,  {Meisel} Z.,  {Smith} K.,
  {Warren} S.,  {Heger} A.,  {Hoffman} R.~D.,  {Rauscher} T.,  {Sakharuk} A.,
  {Schatz} H.,  {Thielemann} F.~K.,    {Wiescher} M.,  2010, \apjs, 189, 240

\bibitem[\protect\citeauthoryear{{D'Antona}, {Bellazzini}, {Caloi}, {Pecci},
  {Galleti} \& {Rood}}{{D'Antona} et~al.}{2005}]{dan05}
{D'Antona} F.,  {Bellazzini} M.,  {Caloi} V.,  {Pecci} F.~F.,  {Galleti} S.,
  {Rood} R.~T.,  2005, \apj, 631, 868

\bibitem[\protect\citeauthoryear{{D'Antona} \& {Caloi}}{{D'Antona} \&
  {Caloi}}{2004}]{dan04}
{D'Antona} F.,  {Caloi} V.,  2004, \apj, 611, 871

\bibitem[\protect\citeauthoryear{{D'Antona}, {D'Ercole}, {Carini}, {Vesperini}
  \& {Ventura}}{{D'Antona} et~al.}{2012}]{dan12}
{D'Antona} F.,  {D'Ercole} A.,  {Carini} R.,  {Vesperini} E.,    {Ventura} P.,
  2012, \mnras, 426, 1710

\bibitem[\protect\citeauthoryear{{D'Antona}, {D'Ercole}, {Marino}, {Milone},
  {Ventura} \& {Vesperini}}{{D'Antona} et~al.}{2011}]{dan11}
{D'Antona} F.,  {D'Ercole} A.,  {Marino} A.~F.,  {Milone} A.~P.,  {Ventura} P.,
     {Vesperini} E.,  2011, \apj, 736, 5

\bibitem[\protect\citeauthoryear{{D'Antona} \& {Ventura}}{{D'Antona} \&
  {Ventura}}{2007}]{dan07}
{D'Antona} F.,  {Ventura} P.,  2007, \mnras, 379, 1431

\bibitem[\protect\citeauthoryear{{de Mink}, {Pols}, {Langer} \& {Izzard}}{{de
  Mink} et~al.}{2009}]{dem09}
{de Mink} S.~E.,  {Pols} O.~R.,  {Langer} N.,    {Izzard} R.~G.,  2009, \aap,
  507, L1

\bibitem[\protect\citeauthoryear{{Decressin}, {Charbonnel}, {Siess},
  {Palacios}, {Meynet} \& {Georgy}}{{Decressin} et~al.}{2009}]{dec09}
{Decressin} T.,  {Charbonnel} C.,  {Siess} L.,  {Palacios} A.,  {Meynet} G.,
  {Georgy} C.,  2009, \aap, 505, 727

\bibitem[\protect\citeauthoryear{{Decressin}, {Meynet}, {Charbonnel},
  {Prantzos} \& {Ekstr{\"o}m}}{{Decressin} et~al.}{2007}]{dec07b}
{Decressin} T.,  {Meynet} G.,  {Charbonnel} C.,  {Prantzos} N.,
  {Ekstr{\"o}m} S.,  2007, \aap, 464, 1029

\bibitem[\protect\citeauthoryear{{Denisenkov} \& {Denisenkova}}{{Denisenkov} \&
  {Denisenkova}}{1990}]{den90}
{Denisenkov} P.~A.,  {Denisenkova} S.~N.,  1990, Soviet Astronomy Letters, 16,
  275

\bibitem[\protect\citeauthoryear{{Denissenkov}, {Da Costa}, {Norris} \&
  {Weiss}}{{Denissenkov} et~al.}{1998}]{den98}
{Denissenkov} P.~A.,  {Da Costa} G.~S.,  {Norris} J.~E.,    {Weiss} A.,  1998,
  \aap, 333, 926

\bibitem[\protect\citeauthoryear{{Denissenkov} \& {Hartwick}}{{Denissenkov} \&
  {Hartwick}}{2014}]{den14}
{Denissenkov} P.~A.,  {Hartwick} F.~D.~A.,  2014, \mnras, 437, L21

\bibitem[\protect\citeauthoryear{{Denissenkov} \& {Herwig}}{{Denissenkov} \&
  {Herwig}}{2003}]{den03}
{Denissenkov} P.~A.,  {Herwig} F.,  2003, \apjl, 590, L99

\bibitem[\protect\citeauthoryear{{D'Ercole}, {D'Antona}, {Carini}, {Vesperini}
  \& {Ventura}}{{D'Ercole} et~al.}{2012}]{der12}
{D'Ercole} A.,  {D'Antona} F.,  {Carini} R.,  {Vesperini} E.,    {Ventura} P.,
  2012, \mnras, 423, 1521

\bibitem[\protect\citeauthoryear{{D'Ercole}, {D'Antona}, {Ventura}, {Vesperini}
  \& {McMillan}}{{D'Ercole} et~al.}{2010}]{der10}
{D'Ercole} A.,  {D'Antona} F.,  {Ventura} P.,  {Vesperini} E.,    {McMillan}
  S.~L.~W.,  2010, \mnras, 407, 854

\bibitem[\protect\citeauthoryear{{D'Ercole}, {Vesperini}, {D'Antona},
  {McMillan} \& {Recchi}}{{D'Ercole} et~al.}{2008}]{der08}
{D'Ercole} A.,  {Vesperini} E.,  {D'Antona} F.,  {McMillan} S.~L.~W.,
  {Recchi} S.,  2008, \mnras, 391, 825

\bibitem[\protect\citeauthoryear{{Doherty}, {Gil-Pons}, {Lau}, {Lattanzio} \&
  {Siess}}{{Doherty} et~al.}{2014}]{doh14a}
{Doherty} C.~L.,  {Gil-Pons} P.,  {Lau} H.~H.~B.,  {Lattanzio} J.~C.,
  {Siess} L.,  2014, \mnras, 437, 195 (Paper II)

\bibitem[\protect\citeauthoryear{{Doherty}, {Siess}, {Lattanzio} \&
  {Gil-Pons}}{{Doherty} et~al.}{2010}]{doh10}
{Doherty} C.~L.,  {Siess} L.,  {Lattanzio} J.~C.,    {Gil-Pons} P.,  2010,
  \mnras, 401, 1453 (Paper I)

\bibitem[\protect\citeauthoryear{{D'Orazi}, {Campbell}, {Lugaro}, {Lattanzio},
  {Pignatari} \& {Carretta}}{{D'Orazi} et~al.}{2013}]{dor13a}
{D'Orazi} V.,  {Campbell} S.~W.,  {Lugaro} M.,  {Lattanzio} J.~C.,  {Pignatari}
  M.,    {Carretta} E.,  2013, \mnras, 433, 366

\bibitem[\protect\citeauthoryear{{D'Orazi} \& {Marino}}{{D'Orazi} \&
  {Marino}}{2010}]{dor10b}
{D'Orazi} V.,  {Marino} A.~F.,  2010, \apjl, 716, L166

\bibitem[\protect\citeauthoryear{{Fenner}, {Campbell}, {Karakas}, {Lattanzio}
  \& {Gibson}}{{Fenner} et~al.}{2004}]{fen04}
{Fenner} Y.,  {Campbell} S.,  {Karakas} A.~I.,  {Lattanzio} J.~C.,    {Gibson}
  B.~K.,  2004, \mnras, 353, 789

\bibitem[\protect\citeauthoryear{{Fenner}, {Gibson}, {Lee}, {Karakas},
  {Lattanzio}, {Chieffi}, {Limongi} \& {Yong}}{{Fenner} et~al.}{2003}]{fen03}
{Fenner} Y.,  {Gibson} B.~K.,  {Lee} H.-c.,  {Karakas} A.~I.,  {Lattanzio}
  J.~C.,  {Chieffi} A.,  {Limongi} M.,    {Yong} D.,  2003, \pasa, 20, 340

\bibitem[\protect\citeauthoryear{{Ferguson}, {Alexander}, {Allard}, {Barman},
  {Bodnarik}, {Hauschildt}, {Heffner-Wong} \& {Tamanai}}{{Ferguson}
  et~al.}{2005}]{fer05b}
{Ferguson} J.~W.,  {Alexander} D.~R.,  {Allard} F.,  {Barman} T.,  {Bodnarik}
  J.~G.,  {Hauschildt} P.~H.,  {Heffner-Wong} A.,    {Tamanai} A.,  2005, \apj,
  623, 585

\bibitem[\protect\citeauthoryear{{Garc{\'{\i}}a-Hern{\'a}ndez},
  {Garc{\'{\i}}a-Lario}, {Plez}, {D'Antona}, {Manchado} \&
  {Trigo-Rodr{\'{\i}}guez}}{{Garc{\'{\i}}a-Hern{\'a}ndez} et~al.}{2006}]{gar06}
{Garc{\'{\i}}a-Hern{\'a}ndez} D.~A.,  {Garc{\'{\i}}a-Lario} P.,  {Plez} B.,
  {D'Antona} F.,  {Manchado} A.,    {Trigo-Rodr{\'{\i}}guez} J.~M.,  2006,
  Science, 314, 1751

\bibitem[\protect\citeauthoryear{{Garc{\'{\i}}a-Hern{\'a}ndez}, {Manchado},
  {Lambert}, {Plez}, {Garc{\'{\i}}a-Lario}, {D'Antona}, {Lugaro}, {Karakas} \&
  {van Raai}}{{Garc{\'{\i}}a-Hern{\'a}ndez} et~al.}{2009}]{gar09}
{Garc{\'{\i}}a-Hern{\'a}ndez} D.~A.,  {Manchado} A.,  {Lambert} D.~L.,  {Plez}
  B.,  {Garc{\'{\i}}a-Lario} P.,  {D'Antona} F.,  {Lugaro} M.,  {Karakas}
  A.~I.,    {van Raai} M.~A.,  2009, \apjl, 705, L31

\bibitem[\protect\citeauthoryear{{Garc{\'{\i}}a-Hern{\'a}ndez}, {Zamora},
  {Yag{\"u}e}, {Uttenthaler}, {Karakas}, {Lugaro}, {Ventura} \&
  {Lambert}}{{Garc{\'{\i}}a-Hern{\'a}ndez} et~al.}{2013}]{gar13}
{Garc{\'{\i}}a-Hern{\'a}ndez} D.~A.,  {Zamora} O.,  {Yag{\"u}e} A.,
  {Uttenthaler} S.,  {Karakas} A.~I.,  {Lugaro} M.,  {Ventura} P.,    {Lambert}
  D.~L.,  2013, \aap, 555, L3

\bibitem[\protect\citeauthoryear{{Gil-Pons}, {Doherty}, {Lau}, {Campbell},
  {Suda}, {Guilani}, {Guti{\'e}rrez} \& {Lattanzio}}{{Gil-Pons}
  et~al.}{2013}]{gil13}
{Gil-Pons} P.,  {Doherty} C.~L.,  {Lau} H.,  {Campbell} S.~W.,  {Suda} T.,
  {Guilani} S.,  {Guti{\'e}rrez} J.,    {Lattanzio} J.~C.,  2013, \aap, 557,
  A106

\bibitem[\protect\citeauthoryear{{Girardi}, {Bressan}, {Chiosi}, {Bertelli} \&
  {Nasi}}{{Girardi} et~al.}{1996}]{gir96}
{Girardi} L.,  {Bressan} A.,  {Chiosi} C.,  {Bertelli} G.,    {Nasi} E.,  1996,
  \aaps, 117, 113

\bibitem[\protect\citeauthoryear{{Gratton}, {Carretta} \&
  {Bragaglia}}{{Gratton} et~al.}{2012}]{gra12}
{Gratton} R.~G.,  {Carretta} E.,    {Bragaglia} A.,  2012, \aapr, 20, 50

\bibitem[\protect\citeauthoryear{{Gratton}, {Carretta}, {Bragaglia},
  {Lucatello} \& {D'Orazi}}{{Gratton} et~al.}{2010}]{gra10}
{Gratton} R.~G.,  {Carretta} E.,  {Bragaglia} A.,  {Lucatello} S.,    {D'Orazi}
  V.,  2010, \aap, 517, A81

\bibitem[\protect\citeauthoryear{{Gratton}, {Lucatello}, {Carretta},
  {Bragaglia}, {D'Orazi} \& {Momany}}{{Gratton} et~al.}{2011}]{gra11}
{Gratton} R.~G.,  {Lucatello} S.,  {Carretta} E.,  {Bragaglia} A.,  {D'Orazi}
  V.,    {Momany} Y.~A.,  2011, \aap, 534, A123

\bibitem[\protect\citeauthoryear{{Grevesse}, {Noels} \& {Sauval}}{{Grevesse}
  et~al.}{1996}]{gre96}
{Grevesse} N.,  {Noels} A.,    {Sauval} A.~J.,  1996, in {Holt} S.~S.,
  {Sonneborn} G.,  eds, Cosmic Abundances Vol.~99 of ASP Conf. Series,
  {Standard Abundances}.
pp 117--+

\bibitem[\protect\citeauthoryear{{Groenewegen}, {Sloan}, {Soszy{\'n}ski} \&
  {Petersen}}{{Groenewegen} et~al.}{2009}]{gro09}
{Groenewegen} M.~A.~T.,  {Sloan} G.~C.,  {Soszy{\'n}ski} I.,    {Petersen}
  E.~A.,  2009, \aap, 506, 1277

\bibitem[\protect\citeauthoryear{{Hale}, {Champagne}, {Iliadis}, {Hansper},
  {Powell} \& {Blackmon}}{{Hale} et~al.}{2002}]{hal02}
{Hale} S.~E.,  {Champagne} A.~E.,  {Iliadis} C.,  {Hansper} V.~Y.,  {Powell}
  D.~C.,    {Blackmon} J.~C.,  2002, \prc, 65, 015801

\bibitem[\protect\citeauthoryear{{Hale}, {Champagne}, {Iliadis}, {Hansper},
  {Powell} \& {Blackmon}}{{Hale} et~al.}{2004}]{hal04}
{Hale} S.~E.,  {Champagne} A.~E.,  {Iliadis} C.,  {Hansper} V.~Y.,  {Powell}
  D.~C.,    {Blackmon} J.~C.,  2004, \prc, 70, 5802

\bibitem[\protect\citeauthoryear{{Harris}}{{Harris}}{1996}]{har96}
{Harris} W.~E.,  1996, \aj, 112, 1487

\bibitem[\protect\citeauthoryear{{Herwig}}{{Herwig}}{2004}]{her04a}
{Herwig} F.,  2004, \apj, 605, 425

\bibitem[\protect\citeauthoryear{{Herwig}, {VandenBerg}, {Navarro}, {Ferguson}
  \& {Paxton}}{{Herwig} et~al.}{2012}]{her12}
{Herwig} F.,  {VandenBerg} D.~A.,  {Navarro} J.~F.,  {Ferguson} J.,    {Paxton}
  B.,  2012, \apj, 757, 132

\bibitem[\protect\citeauthoryear{{Iben}, {Ritossa} \& {Garcia-Berro}}{{Iben}
  et~al.}{1997}]{pap4}
{Iben} I.~J.,  {Ritossa} C.,    {Garcia-Berro} E.,  1997, \apj, 489, 772

\bibitem[\protect\citeauthoryear{{Iglesias} \& {Rogers}}{{Iglesias} \&
  {Rogers}}{1996}]{opal}
{Iglesias} C.~A.,  {Rogers} F.~J.,  1996, \apj, 464, 943

\bibitem[\protect\citeauthoryear{{Iliadis}, {D'Auria}, {Starrfield}, {Thompson}
  \& {Wiescher}}{{Iliadis} et~al.}{2001}]{ili01}
{Iliadis} C.,  {D'Auria} J.~M.,  {Starrfield} S.,  {Thompson} W.~J.,
  {Wiescher} M.,  2001, \apjs, 134, 151

\bibitem[\protect\citeauthoryear{{Iliadis}, {Longland}, {Champagne}, {Coc} \&
  {Fitzgerald}}{{Iliadis} et~al.}{2010}]{ili10}
{Iliadis} C.,  {Longland} R.,  {Champagne} A.~E.,  {Coc} A.,    {Fitzgerald}
  R.,  2010, Nuclear Physics A, 841, 31

\bibitem[\protect\citeauthoryear{{Ivans}, {Sneden}, {Kraft}, {Suntzeff},
  {Smith}, {Langer} \& {Fulbright}}{{Ivans} et~al.}{1999}]{iva99}
{Ivans} I.~I.,  {Sneden} C.,  {Kraft} R.~P.,  {Suntzeff} N.~B.,  {Smith} V.~V.,
   {Langer} G.~E.,    {Fulbright} J.~P.,  1999, \aj, 118, 1273

\bibitem[\protect\citeauthoryear{{Izzard}, {Lugaro}, {Karakas}, {Iliadis} \&
  {van Raai}}{{Izzard} et~al.}{2007}]{izz07}
{Izzard} R.~G.,  {Lugaro} M.,  {Karakas} A.~I.,  {Iliadis} C.,    {van Raai}
  M.,  2007, \aap, 466, 641

\bibitem[\protect\citeauthoryear{{Jorissen} \& {Arnould}}{{Jorissen} \&
  {Arnould}}{1989}]{jor89}
{Jorissen} A.,  {Arnould} M.,  1989, \aap, 221, 161

\bibitem[\protect\citeauthoryear{{Karakas}}{{Karakas}}{2010}]{kar10}
{Karakas} A.~I.,  2010, \mnras, 403, 1413

\bibitem[\protect\citeauthoryear{{Karakas}, {Lugaro}, {Wiescher}, {G{\"o}rres}
  \& {Ugalde}}{{Karakas} et~al.}{2006}]{kar06a}
{Karakas} A.~I.,  {Lugaro} M.~A.,  {Wiescher} M.,  {G{\"o}rres} J.,    {Ugalde}
  C.,  2006, \apj, 643, 471

\bibitem[\protect\citeauthoryear{{Kobayashi}, {Umeda}, {Nomoto}, {Tominaga} \&
  {Ohkubo}}{{Kobayashi} et~al.}{2006}]{kob06}
{Kobayashi} C.,  {Umeda} H.,  {Nomoto} K.,  {Tominaga} N.,    {Ohkubo} T.,
  2006, \apj, 653, 1145

\bibitem[\protect\citeauthoryear{{Kraft}}{{Kraft}}{1979}]{kra79}
{Kraft} R.~P.,  1979, \araa, 17, 309

\bibitem[\protect\citeauthoryear{{Kraft}}{{Kraft}}{1994}]{kra94}
{Kraft} R.~P.,  1994, \pasp, 106, 553

\bibitem[\protect\citeauthoryear{{Kroupa}, {Tout} \& {Gilmore}}{{Kroupa}
  et~al.}{1993}]{kro93}
{Kroupa} P.,  {Tout} C.~A.,    {Gilmore} G.,  1993, \mnras, 262, 545

\bibitem[\protect\citeauthoryear{{Lagadec} \& {Zijlstra}}{{Lagadec} \&
  {Zijlstra}}{2008}]{lag08}
{Lagadec} E.,  {Zijlstra} A.~A.,  2008, \mnras, 390, L59

\bibitem[\protect\citeauthoryear{{Lattanzio}}{{Lattanzio}}{1986}]{lat86}
{Lattanzio} J.~C.,  1986, \apj, 311, 708

\bibitem[\protect\citeauthoryear{{Lau}, {Gil-Pons}, {Doherty} \&
  {Lattanzio}}{{Lau} et~al.}{2012}]{lau12}
{Lau} H.~H.~B.,  {Gil-Pons} P.,  {Doherty} C.,    {Lattanzio} J.,  2012, \aap,
  542, A1

\bibitem[\protect\citeauthoryear{{Lederer} \& {Aringer}}{{Lederer} \&
  {Aringer}}{2009}]{led09}
{Lederer} M.~T.,  {Aringer} B.,  2009, \aap, 494, 403

\bibitem[\protect\citeauthoryear{{Lee}, {Joo}, {Han}, {Chung}, {Ree}, {Sohn},
  {Kim}, {Yoon}, {Yi} \& {Demarque}}{{Lee} et~al.}{2005}]{lee05}
{Lee} Y.-W.,  {Joo} S.-J.,  {Han} S.-I.,  {Chung} C.,  {Ree} C.~H.,  {Sohn}
  Y.-J.,  {Kim} Y.-C.,  {Yoon} S.-J.,  {Yi} S.~K.,    {Demarque} P.,  2005,
  \apjl, 621, L57

\bibitem[\protect\citeauthoryear{{Longland}, {Iliadis} \& {Karakas}}{{Longland}
  et~al.}{2012}]{lon12}
{Longland} R.,  {Iliadis} C.,    {Karakas} A.~I.,  2012, \prc, 85, 065809

\bibitem[\protect\citeauthoryear{{Lugaro}, {Karakas}, {Stancliffe} \&
  {Rijs}}{{Lugaro} et~al.}{2012}]{lug12a}
{Lugaro} M.,  {Karakas} A.~I.,  {Stancliffe} R.~J.,    {Rijs} C.,  2012, \apj,
  747, 2

\bibitem[\protect\citeauthoryear{{Lugaro}, {Ugalde}, {Karakas}, {G{\"o}rres},
  {Wiescher}, {Lattanzio} \& {Cannon}}{{Lugaro} et~al.}{2004}]{lug04}
{Lugaro} M.,  {Ugalde} C.,  {Karakas} A.~I.,  {G{\"o}rres} J.,  {Wiescher} M.,
  {Lattanzio} J.~C.,    {Cannon} R.~C.,  2004, \apj, 615, 934

\bibitem[\protect\citeauthoryear{{Marigo}}{{Marigo}}{2002}]{mar02}
{Marigo} P.,  2002, \aap, 387, 507

\bibitem[\protect\citeauthoryear{{Marigo} \& {Aringer}}{{Marigo} \&
  {Aringer}}{2009}]{mar09}
{Marigo} P.,  {Aringer} B.,  2009, \aap, 508, 1539

\bibitem[\protect\citeauthoryear{{Marino}, {Milone}, {Przybilla}, {Bergemann},
  {Lind}, {Asplund}, {Cassisi}, {Catelan}, {Casagrande}, {Valcarce}, {Bedin},
  {Cortes}, {D'Antona}, {Jerjen}, {Piotto}, {Schlesinger}, {Zoccali} \&
  {Angeloni}}{{Marino} et~al.}{2013}]{mar13}
{Marino} A.~F.,  {Milone} A.~P.,  {Przybilla} N.,  {Bergemann} M.,  {Lind} K.,
  {Asplund} M.,  {Cassisi} S.,  {Catelan} M.,  {Casagrande} L.,  {Valcarce}
  A.~A.~R.,  {Bedin} L.~R.,  {Cortes} C.,  {D'Antona} F.,  {Jerjen} H.,
  {Piotto} G.,  {Schlesinger} K.,  {Zoccali} M.,    {Angeloni} R.,  2013, ArXiv
  e-prints

\bibitem[\protect\citeauthoryear{{Mattsson}, {Wahlin}, {H{\"o}fner} \&
  {Eriksson}}{{Mattsson} et~al.}{2008}]{mat08b}
{Mattsson} L.,  {Wahlin} R.,  {H{\"o}fner} S.,    {Eriksson} K.,  2008, \aap,
  484, L5

\bibitem[\protect\citeauthoryear{{McSaveney}, {Wood}, {Scholz}, {Lattanzio} \&
  {Hinkle}}{{McSaveney} et~al.}{2007}]{mcs07}
{McSaveney} J.~A.,  {Wood} P.~R.,  {Scholz} M.,  {Lattanzio} J.~C.,    {Hinkle}
  K.~H.,  2007, \mnras, 378, 1089

\bibitem[\protect\citeauthoryear{{Norris}}{{Norris}}{2004}]{nor04}
{Norris} J.~E.,  2004, \apjl, 612, L25

\bibitem[\protect\citeauthoryear{{Pasquini}, {Mauas}, {K{\"a}ufl} \&
  {Cacciari}}{{Pasquini} et~al.}{2011}]{pas11}
{Pasquini} L.,  {Mauas} P.,  {K{\"a}ufl} H.~U.,    {Cacciari} C.,  2011, \aap,
  531, A35

\bibitem[\protect\citeauthoryear{{Piotto}, {Bedin}, {Anderson}, {King},
  {Cassisi}, {Milone}, {Villanova}, {Pietrinferni} \& {Renzini}}{{Piotto}
  et~al.}{2007}]{pio07}
{Piotto} G.,  {Bedin} L.~R.,  {Anderson} J.,  {King} I.~R.,  {Cassisi} S.,
  {Milone} A.~P.,  {Villanova} S.,  {Pietrinferni} A.,    {Renzini} A.,  2007,
  \apjl, 661, L53

\bibitem[\protect\citeauthoryear{{Pumo}, {D'Antona} \& {Ventura}}{{Pumo}
  et~al.}{2008}]{pum08}
{Pumo} M.~L.,  {D'Antona} F.,    {Ventura} P.,  2008, \apjl, 672, L25

\bibitem[\protect\citeauthoryear{{Reimers}}{{Reimers}}{1975}]{rei75}
{Reimers} D.,  1975, Memoires of the Societe Royale des Sciences de Liege, 8,
  369

\bibitem[\protect\citeauthoryear{{Sackmann}}{{Sackmann}}{1977}]{sac77}
{Sackmann} I.-J.,  1977, \apj, 212, 159

\bibitem[\protect\citeauthoryear{{Siess}}{{Siess}}{2006}]{sie06}
{Siess} L.,  2006, \aap, 448, 717

\bibitem[\protect\citeauthoryear{{Siess}}{{Siess}}{2010}]{sie10}
{Siess} L.,  2010, \aap, 512, A10+

\bibitem[\protect\citeauthoryear{{Siess} \& {Arnould}}{{Siess} \&
  {Arnould}}{2008}]{sie08b}
{Siess} L.,  {Arnould} M.,  2008, \aap, 489, 395

\bibitem[\protect\citeauthoryear{{Smith}, {Shetrone}, {Bell}, {Churchill} \&
  {Briley}}{{Smith} et~al.}{1996}]{smi96}
{Smith} G.~H.,  {Shetrone} M.~D.,  {Bell} R.~A.,  {Churchill} C.~W.,
  {Briley} M.~M.,  1996, \aj, 112, 1511

\bibitem[\protect\citeauthoryear{{Straniero}, {Gallino} \&
  {Cristallo}}{{Straniero} et~al.}{2006}]{str06}
{Straniero} O.,  {Gallino} R.,    {Cristallo} S.,  2006, Nuclear Physics A,
  777, 311

\bibitem[\protect\citeauthoryear{{Straniero}, {Imbriani}, {Strieder},
  {Bemmerer}, {Broggini}, {Caciolli}, {Corvisiero}, {Costantini}, {et} \&
  {al}}{{Straniero} et~al.}{2013}]{str13}
{Straniero} O.,  {Imbriani} G.,  {Strieder} F.,  {Bemmerer} D.,  {Broggini} C.,
   {Caciolli} A.,  {Corvisiero} P.,  {Costantini} H.,  {et}   {al} 2013, \apj,
  763, 100

\bibitem[\protect\citeauthoryear{{van Loon}, {Zijlstra} \& {Groenewegen}}{{van
  Loon} et~al.}{1999}]{van99}
{van Loon} J.~T.,  {Zijlstra} A.~A.,    {Groenewegen} M.~A.~T.,  1999, \aap,
  346, 805

\bibitem[\protect\citeauthoryear{{van Raai}, {Lugaro}, {Karakas},
  {Garc{\'{\i}}a-Hern{\'a}ndez} \& {Yong}}{{van Raai} et~al.}{2012}]{van12}
{van Raai} M.~A.,  {Lugaro} M.,  {Karakas} A.~I.,
  {Garc{\'{\i}}a-Hern{\'a}ndez} D.~A.,    {Yong} D.,  2012, \aap, 540, A44

\bibitem[\protect\citeauthoryear{{van Raai}, {Lugaro}, {Karakas} \&
  {Iliadis}}{{van Raai} et~al.}{2008}]{van08}
{van Raai} M.~A.,  {Lugaro} M.,  {Karakas} A.~I.,    {Iliadis} C.,  2008, \aap,
  478, 521

\bibitem[\protect\citeauthoryear{{Vassiliadis} \& {Wood}}{{Vassiliadis} \&
  {Wood}}{1993}]{vas93}
{Vassiliadis} E.,  {Wood} P.~R.,  1993, \apj, 413, 641

\bibitem[\protect\citeauthoryear{{Ventura}, {Carini} \& {D'Antona}}{{Ventura}
  et~al.}{2011}]{ven11b}
{Ventura} P.,  {Carini} R.,    {D'Antona} F.,  2011, \mnras, 415, 3865

\bibitem[\protect\citeauthoryear{{Ventura} \& {D'Antona}}{{Ventura} \&
  {D'Antona}}{2005}]{ven05}
{Ventura} P.,  {D'Antona} F.,  2005, \aap, 431, 279

\bibitem[\protect\citeauthoryear{{Ventura} \& {D'Antona}}{{Ventura} \&
  {D'Antona}}{2009}]{ven09b}
{Ventura} P.,  {D'Antona} F.,  2009, \aap, 499, 835

\bibitem[\protect\citeauthoryear{{Ventura} \& {D'Antona}}{{Ventura} \&
  {D'Antona}}{2010}]{ven10a}
{Ventura} P.,  {D'Antona} F.,  2010, \mnras, 402, L72

\bibitem[\protect\citeauthoryear{{Ventura}, {D'Antona} \&
  {Mazzitelli}}{{Ventura} et~al.}{2000}]{ven00}
{Ventura} P.,  {D'Antona} F.,    {Mazzitelli} I.,  2000, \aap, 363, 605

\bibitem[\protect\citeauthoryear{{Ventura}, {D'Antona} \&
  {Mazzitelli}}{{Ventura} et~al.}{2002}]{ven02}
{Ventura} P.,  {D'Antona} F.,    {Mazzitelli} I.,  2002, \aap, 393, 215

\bibitem[\protect\citeauthoryear{{Ventura}, {D'Antona}, {Mazzitelli} \&
  {Gratton}}{{Ventura} et~al.}{2001}]{ven01}
{Ventura} P.,  {D'Antona} F.,  {Mazzitelli} I.,    {Gratton} R.,  2001, \apjl,
  550, L65

\bibitem[\protect\citeauthoryear{{Ventura}, {Di Criscienzo}, {Carini} \&
  {D'Antona}}{{Ventura} et~al.}{2013}]{ven13}
{Ventura} P.,  {Di Criscienzo} M.,  {Carini} R.,    {D'Antona} F.,  2013,
  \mnras, 431, 3642

\bibitem[\protect\citeauthoryear{{Ventura} \& {Marigo}}{{Ventura} \&
  {Marigo}}{2010}]{ven10b}
{Ventura} P.,  {Marigo} P.,  2010, \mnras, 408, 2476

\bibitem[\protect\citeauthoryear{{Villanova}, {Geisler} \&
  {Piotto}}{{Villanova} et~al.}{2010}]{vil10}
{Villanova} S.,  {Geisler} D.,    {Piotto} G.,  2010, \apjl, 722, L18

\bibitem[\protect\citeauthoryear{{Wachter}, {Winters}, {Schr{\"o}der} \&
  {Sedlmayr}}{{Wachter} et~al.}{2008}]{wac08}
{Wachter} A.,  {Winters} J.~M.,  {Schr{\"o}der} K.-P.,    {Sedlmayr} E.,  2008,
  \aap, 486, 497

\bibitem[\protect\citeauthoryear{{Wood} \& {Arnett}}{{Wood} \&
  {Arnett}}{2011}]{woo11a}
{Wood} P.,  {Arnett} D.,  2011 Vol.~445 of ASP Conf. Series, {Testing a
  Modified Mixing-Length Theory: Comparison to the Pulsation of AGB Stars}.
p.~183

\bibitem[\protect\citeauthoryear{{Wood}}{{Wood}}{1981}]{woo81c}
{Wood} P.~R.,  1981, in {Iben} Jr. I.,  {Renzini} A.,  eds, Physical Processes
  in Red Giants Vol.~88 of Astrophysics and Space Science Library, {The
  conditions for dredge-up of carbon during the helium shell flash and the
  production of carbon stars}.
pp 135--139

\bibitem[\protect\citeauthoryear{{Wood} \& {Faulkner}}{{Wood} \&
  {Faulkner}}{1986}]{woo86}
{Wood} P.~R.,  {Faulkner} D.~J.,  1986, \apj, 307, 659

\bibitem[\protect\citeauthoryear{{Yong}, {Aoki}, {Lambert} \& {Paulson}}{{Yong}
  et~al.}{2006}]{yon06b}
{Yong} D.,  {Aoki} W.,  {Lambert} D.~L.,    {Paulson} D.~B.,  2006, \apj, 639,
  918

\bibitem[\protect\citeauthoryear{{Yong}, {Grundahl}, {D'Antona}, {Karakas},
  {Lattanzio} \& {Norris}}{{Yong} et~al.}{2009}]{yon09}
{Yong} D.,  {Grundahl} F.,  {D'Antona} F.,  {Karakas} A.~I.,  {Lattanzio}
  J.~C.,    {Norris} J.~E.,  2009, \apjl, 695, L62

\bibitem[\protect\citeauthoryear{{Yong}, {Lambert}, {Paulson} \&
  {Carney}}{{Yong} et~al.}{2008}]{yon08a}
{Yong} D.,  {Lambert} D.~L.,  {Paulson} D.~B.,    {Carney} B.~W.,  2008, \apj,
  673, 854

\end{thebibliography}
\appendix

\section{Supplementary Tables}
\begin{table*}
\begin{center} 
\caption{Initial and (corrosive) second dredge-up surface abundances of selected isotopes for our standard VW93 models for metallicities Z=0.001 and 0.0001. Where $n(m)= n\times 10^{m}$.}
\label{initial2du} \setlength{\tabcolsep}{2pt} 
\begin{tabular}{lrcccccccccccccc} \hline
$M_{\rm{ini}}$& Phase&\chem{1}H&\chem{4}He&\chem{12}C&\chem{13}C&\chem{14}N&\chem{15}N&\chem{16}O&\chem{17}O&\chem{18}O&\chem{21}Ne&\chem{22}Ne&\chem{23}Na&\chem{25}Mg&\chem{26}Mg\\ \hline
\hline \multicolumn{16}{c}{Z=0.001} \\ \hline
- &Initial&0.7498 &0.2492 &1.82(-4)&2.05(-6)&5.58(-5)&2.07(-7)&5.06(-4)&1.94(-7)&1.08(-6)&2.31(-7)&7.27(-6)&1.99(-6)&3.85(-6)& 4.40(-6)\\ \hline
6.5 &2DU  &0.6501 &0.3487 &1.37(-4)&4.69(-6)&2.40(-4)&6.43(-8)&4.21(-4)&2.60(-6)&1.72(-6)&2.69(-7)&7.10(-6)&8.85(-6)&3.23(-6)& 4.40(-6)\\
7.0   &C2DU  &0.6488 &0.3500 &1.95(-4)&6.18(-6)&2.44(-4)&3.82(-8)&4.20(-4)&2.56(-6)&1.87(-6)&2.81(-7)&8.14(-6)&8.90(-6)&3.20(-6)& 4.37(-6)\\
7.5 &CO2DU  &0.6449 &0.3525 &1.20(-3)&5.38(-5)&3.07(-4)&3.98(-8)&7.05(-4)&2.42(-6)&1.44(-6)&2.90(-7)&1.33(-5)&9.06(-6)&3.20(-6)&4.37(-6)\\ 
\hline \multicolumn{16}{c}{Z=0.0001} \\ \hline
- &Initial&0.7519 &0.2480 &1.82(-5)&2.05(-7)&5.58(-6)&2.07(-8)&5.06(-5)&1.94(-8)&1.08(-7)&2.31(-8)&7.27(-7)&1.99(-7)&3.85(-7)& 4.40(-7)\\ \hline
6.5 &C2DU &0.6505  &0.3493 &6.52(-5)& 4.07(-7)& 2.89(-5)& 4.96(-9) & 3.95(-5)&4.90(-7)&1.63(-7)&3.35(-8)&6.58(-7)&9.32(-7)& 3.13(-7)& 3.97(-7)\\ 
7.0 &C2DU   &0.6492  &0.3506 &1.65(-4)&1.62(-6)&2.99(-5)&2.68(-9)&4.14(-5)&4.53(-7)&1.73(-7)&3.30(-8)&7.95(-7)& 9.26(-7)&3.09(-7)&3.99(-7) \\ 
7.5 &CO2DU & 0.6461 & 0.3520 & 1.29(-3) & 3.40(-5) & 4.96(-5) & 3.49(-9) & 4.20(-4) & 4.61(-7) & 1.46(-7) & 3.61(-8) & 1.24(-6) & 9.49(-7) & 3.14(-7) & 4.01(-7) \\ 
\hline 
\end{tabular}
 \end{center}
 \end{table*}

\begin{table*}
\begin{center}\setlength{\tabcolsep}{3pt} 
\caption{Sample first few lines of yield table. The production factor is zero for species whose initial composition is also zero (e.g. \chem{26}Al).}\label{tableappend2}
\begin{tabular}{ccccccc}
\hline
\multicolumn{7}{c}{$M_{\rm{ini}}$ = 6.5\,M$_\odot$ Z = 0.001} \\ 
&&&&\multicolumn{3}{c}{Extrapolated TPs}\\ \cline{5-7}
Species $i$ & Net yield (M$_{\odot}$) & $M^{\mathrm{wind}}(i)$ (M$_{\odot}$) &$\log_{10}[\langle X(i)
  \rangle/X_{\mathrm{ini}}(i)]$ & Net yield (M$_{\odot}$) &$M^{\mathrm{wind}}(i)$ (M$_{\odot}$) &$\log_{10}[\langle X(i)\rangle/X_{\mathrm{ini}}(i)]$ \\
\hline
     H     & -6.253E-01 & 3.450E+00  &-7.236E-02  & -6.297E-01 & 3.446E+00 & -7.292E-02  \\
\chem{3}He &  1.767E-07 & 1.767E-07  & 0.000E+00  & 1.767E-07  & 1.767E-07 &  0.000E+00 \\
\chem{4}He &  5.970E-01 & 1.951E+00  & 1.586E-01  & 5.991E-01  & 1.954E+00 &  1.590E-01 \\
\chem{7}Li & -5.638E-09 & 6.944E-11  & -1.915E+00 & -5.638E-09 & 6.944E-11 & -1.915E+00 \\
\hline
\end{tabular}
\medskip\\
\end{center}
\end{table*}

\begin{table*}
\begin{center}
\setlength{\tabcolsep}{3pt} 
\caption{Selected model characteristics for the thermal pulsing (S)AGB phase including extrapolated thermal pulses. Here $L^{\rm{Max}}$ is the maximum quiescent luminosity during the TP-(S)AGB phase, C/O$_{\rm{2DU}}$ and C/O$_{\rm{F}}$ are the number ratios at the completion of 2DU and at the end of the evolution respectively, and ($\tau_{\rm{C}}$/$\tau_{\rm{M}}$) is the ratio of carbon rich to oxygen rich duration of the TP-(S)AGB phase. Models denoted with * are those in which no extrapolated TPs were required, however, we have synthetically modelled the resultant 3DU from the last computed TP which ceased to converge prior to completion of 3DU.}\label{table3}
\begin{tabular}{lcccrccccrccc}
 \hline 
 & & & & && & & \multicolumn{5}{c}{Extrapolated TPs}\\ \cline{9-13}
$M_{\rm{ini}}$ & C/O$_{\rm{2DU}}$ &$L^{\rm{Max}}$&  $M_{\rm{Dredge}}^{\rm{Tot}}$ & $N_{TP}$ & $\tau_{\rm{(S)AGB}}$ & C/O$_{\rm{F}}$ & $\tau_{\rm{C}}$/$\tau_{\rm{M}}$ & $M_{\rm{Dredge}}^{\rm{Tot}}$ &  $N_{TP}$& $\tau_{\rm{(S)AGB}}$  & C/O$_{\rm{F}}$ & $\tau_{\rm{C}}$/$\tau_{\rm{M}}$ \\
 (M$_{\odot}$)&   &(L$_{\odot}$)&(M$_{\odot}$)&& (yrs)    &    &   &(M$_{\odot}$)&& (yrs) & & \\ 
\hline
\multicolumn{13}{c}{Z = 0.001}  \\ \hline
6.5         &0.44&8.20E+04&8.32E-02&118&1.69E+05 &12.00&0.63&8.92E-02&123&1.79E+05&21.47 &0.65 \\ 
6.5B        &0.44&6.64E+04&2.01E-02&38 &4.78E+04 &3.83 &0.17&2.11E-02&*38&4.78E+04&6.34  &0.17 \\
7.0         &0.63&9.50E+04&3.97E-02&126&7.83E+04 &9.69 &0.49&4.43E-02&138&8.79E+04&19.39 &0.58 \\ 
7.0B        &0.65&8.42E+04&1.07E-02&47 &2.45E+04 &4.03 &0.16&1.11E-02&*47&2.45E+04&5.53  &0.16 \\ 
7.0B$\alpha$&0.66&9.89E+04&4.37E-03&28 &1.23E+04 &4.56 &0.16&4.63E-03&*28&1.23E+04&8.46  &0.16 \\ 
7.5         &2.35&1.10E+05&2.64E-02&171&5.22E+04 &3.61 &0.21&3.25E-02&197&6.38E+04&13.05 &0.36 \\ 
7.5B        &2.51&1.04E+05&5.81E-03&56 &1.43E+04 &1.78 &0.06&6.17E-03&57 &1.50E+04&2.64  &0.10 \\ 
\hline
\multicolumn{13}{c}{Z = 0.0001}  \\ \hline
6.5         &2.18&1.14E+05&2.72E-01&439&5.00E+05 &2.82 &0.96&2.74E-01&441&5.03E+05&9.54  &0.96 \\ 
6.5O        &2.27&9.86E+04&1.28E-01&214&2.55E+05 &9.07 &0.93&1.33E-01&221&2.64E+05&17.55 &0.94 \\ 6.5B        &2.26&7.78E+04&1.46E-02&35 &3.68E+04 &10.28&1.00&1.54E-02&*35&3.68E+05&21.64 &1.00 \\ 
7.0         &5.29&1.29E+05&1.23E-01&459&2.24E+05 &3.64 &0.97&1.27E-01&470&2.31E+05&11.42 &0.97 \\ 
7.0O        &5.31&1.17E+05&6.30E-02&269&1.26E+05 &8.23 &0.90&6.74E-02&283&1.35E+05&16.02 &0.90 \\ 
7.0B        &5.33&9.76E+04&5.88E-03& 38&1.58E+04 &12.87&0.82&6.20E-03&*38&1.58E+04&18.75 &0.82 \\ 
7.5         &4.19&1.47E+05&5.41E-02&445&1.06E+05 &6.36 &0.71&5.73E-02&464&1.12E+05&12.73 &0.73 \\ 
7.5O        &4.40&1.32E+05&3.03E-02&248&6.23E+04 &6.76 &0.52&3.49E-02&279&7.32E+04&13.76 &0.59 \\ 7.5B        &4.52&1.23E+05&2.30E-03&35 &6.82E+03 &0.39 &0.68&2.53E-03& 36&7.14E+03&2.38  &0.65 \\ 
\hline 
\end{tabular}
\medskip\\
\end{center}
\end{table*}

\section{Online material} We include supplementary electronic online tables as follows\\
{\bf{Table~1.}} Stellar Yields for Z=0.001 and 0.0001 with VW93 mass-loss rate.\\
{\bf{Table~2.}} Stellar Yields for Z=0.001 and 0.0001 with B95 mass-loss rate.\\
{\bf{Table~3.}} Stellar Yields for Z=0.0001 with updated opacities.\\
{\bf{Table~4.}} Initial Composition in mass fraction for Z=0.001 and 0.0001.\\
\end{document}